%% file: ANA-HIGG-2017-13-PAPER.tex
\documentclass[cernpreprint, texlive=2016, UKenglish, texmf]{atlasdoc}

\usepackage{atlaspackage}
 
\usepackage{atlasbiblatex}
\usepackage{atlasprocess}
\usepackage{atlasphysics}
 
\graphicspath{{logos/}{figures/}}
 
\usepackage{ANA-HIGG-2017-13-PAPER-defs}
\usepackage{multirow}

\AtlasTitle{Constraints on Higgs boson properties using $WW^{*}(\rightarrow e\nu\mu\nu)jj$ production in
$\SI{36.1}{\per\femto\barn}$ of $\sqrt{s}=13$~TeV $pp$ collisions with the ATLAS detector}
 
\AtlasAbstract{
This article presents the results of two studies of Higgs boson properties using the
$WW^*(\rightarrow e\nu\mu\nu)jj$ final state, based on a dataset corresponding
to $\SI{36.1}{\per\femto\barn}$ of $\sqrt{s}=13$~TeV proton--proton collisions recorded by the ATLAS experiment
at the Large Hadron Collider. The first study targets Higgs boson production via gluon--gluon fusion and constrains
the CP properties of the effective Higgs--gluon interaction. Using angular distributions and the overall rate, a value
of $\tan(\alpha) = 0.0 \pm 0.4 (\mathrm{stat.}) \pm 0.3 (\mathrm{syst.})$ is obtained for the tangent of the mixing angle
for CP-even and CP-odd contributions. The second study exploits the vector-boson fusion production mechanism to probe the
Higgs boson couplings to longitudinally and transversely polarised $W$ and $Z$ bosons
in both the production and the decay of the Higgs boson; these couplings have not been directly constrained previously.
The polarisation-dependent coupling-strength scale factors are defined as the ratios of the measured polarisation-dependent coupling strengths
to those predicted by the Standard Model, and are determined using rate and kinematic information to be
$a_\mathrm{L}=0.91^{+0.10}_{-0.18}$(stat.)$^{+0.09}_{-0.17}$(syst.) and  $a_{\mathrm{T}}=1.2 \pm 0.4 $(stat.)$ ^{+0.2}_{-0.3} $(syst.).
These coupling strengths are translated into pseudo-observables, resulting in
$\kappa_{VV}= 0.91^{+0.10}_{-0.18}$(stat.)$^{+0.09}_{-0.17}$(syst.) and $\epsilon_{VV} =0.13^{+0.28}_{-0.20}$ (stat.)$^{+0.08}_{-0.10}$(syst.).
All results are consistent with the Standard Model predictions.
}

\author{The ATLAS Collaboration}
 
\AtlasRefCode{HIGG-2017-13}
 
\AtlasJournalRef{Eur. Phys. J. C 82 (2022) 622}
 
\AtlasDOI{10.1140/epjc/s10052-022-10366-1}

\AtlasJournal{Eur.\ Phys.\ J. C}
\PreprintIdNumber{CERN-EP-2021-096}

\hypersetup{pdftitle={ATLAS document},pdfauthor={The ATLAS Collaboration}}
 
\begin{document}
 
\title{\AtlasTitleText}
 
\author{The ATLAS Collaboration}

\maketitle

\section{Introduction}\label{sec:intro}
 
The Higgs boson ($H$) discovery at the Large Hadron Collider (LHC) was a great success of the ATLAS
and CMS Collaborations \cite{Aad:2012tfa,Chatrchyan:2012xd}. Since the discovery, multiple
studies have been performed to establish whether the Higgs boson has the properties predicted by the Standard Model (SM), or instead
is a particle of an as yet unobserved sector beyond the SM.
The mass of the Higgs boson has been measured by the ATLAS and CMS experiments to be $m_H = 125.09 \pm 0.21(\mathrm{stat.})\pm 0.11 (\mathrm{syst.}) \;\GeV$ \cite{Aad:2015zhl}
using $\sqrt{s}=7$ and $8\;\TeV$ $pp$ collision data,
and there are strong indications that the spin and parity states are $J^{P}=0^{+}$ \cite{Aad:2013xqa,Aad:2015mxa,Chatrchyan:2012jja}.
By probing the final-state particles in $H\rightarrow WW^{*}$, $H\rightarrow ZZ^{*}$, and $H\rightarrow \gamma\gamma$ decays, a pure CP-odd
Higgs boson has been excluded at a confidence level above $99.9\%$. In addition, a potential CP-odd
contribution to the $HWW$ and $HZZ$ couplings has been significantly constrained \cite{ATLAS-CONF-2019-050,HVVconstraintsFromCMS,Khachatryan:2016tnr}.
Recently, studies of the CP properties of the top-quark Yukawa coupling using $Htt$ events, and the $\tau$-lepton Yukawa coupling using
$H\rightarrow\tau\tau$ decays have been published \cite{PhysRevLett.125.061802,Sirunyan:2020sum,CMS:2021sdq}.\par
 
The gluon--gluon fusion ($\mathrm{ggF}$) process probes a different kinematic regime than $Htt$ production, and the properties of the effective $ggH$ interaction
itself may differ from SM predictions if there is a loop contribution from previously unobserved particles. The CP nature of the effective coupling between the Higgs
boson and gluons, $ggH$, in the gluon--gluon fusion production mode \cite{Dolan:2014upa} has been studied in Refs.~\cite{Aaboud:2018xdt, Sirunyan:2021fpv, Aad:2015tna}.
This article presents new analysis techniques for enhancing the sensitivity to interference effects between the CP-even and CP-odd contributions to the
Higgs--gluon coupling, and provides constraints on the mixing angle between CP-even and CP-odd contributions. The approach is therefore complementary to
those used in previous studies of the CP properties of the Higgs boson.\par
 
The vector-boson fusion (VBF) Higgs boson production process has been measured by ATLAS and CMS in
numerous channels \cite{Sirunyan:2018egh,Sirunyan:2019nbs,Sirunyan:2019twz,ATLAS-CONF-2019-050,Aaboud:2018gay,atlas-higg-2016-07}.
While measurements of $\sigma_\mathrm{VBF} \cdot \mathcal{B}_{H\to WW}$ are consistent with the SM,
individual polarisation-dependent Higgs boson couplings to the massive electroweak gauge bosons have so far not been studied directly.
Longitudinally polarised electroweak bosons emerge from massless degrees of freedom of the Higgs boson and are therefore closely related to the mechanism of
electroweak symmetry breaking.
The strength of the Higgs boson coupling to longitudinally polarised $W$ bosons ensures the unitarity of the SM.
However, if the Higgs field is not associated with a fundamental scalar particle but is an effective field arising from new dynamics, the coupling may deviate from its SM value.
For example, Higgs compositeness models~\cite{Kaplan:1983fs,KAPLAN1984187} predict more degrees of freedom, allowing
the Higgs boson couplings to electroweak bosons to deviate from their SM values.
 
\begin{figure}[!ht]
\centering
\subfloat[]{\includegraphics[width=0.4\textwidth]{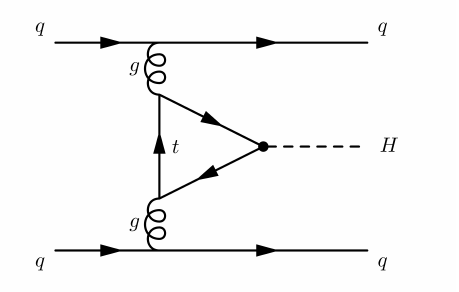}}
\subfloat[]{\includegraphics[width=0.4\textwidth]{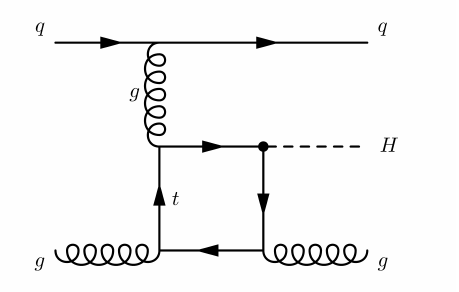}}\\
\subfloat[]{\includegraphics[width=0.4\textwidth]{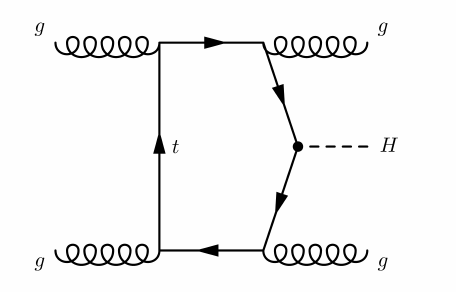}}
\subfloat[]{\includegraphics[width=0.4\textwidth]{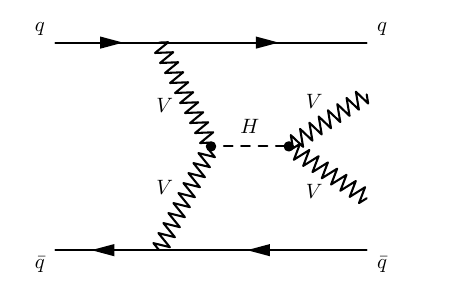}}
\caption{Examples of Feynman diagrams contributing to the production of a Higgs boson in
association with two jets via the fusion of two gluons or two vector bosons ($V$) at leading order in QCD.
The presented diagrams show examples for the subprocesses (a) $qq\rightarrow H qq$, (b)
$qg\rightarrow H qg$ and (c) $gg\rightarrow H gg$, as well as (d) the vector-boson fusion process and
the subsequent decay of the Higgs boson into two vector bosons.}
\label{fig:FeynmandiagramsGGF}
\end{figure}

This article presents the results of two analyses studying the properties of the Higgs boson using its decay into
$WW^{*}\rightarrow e\nu\mu\nu$ and its production in association with two jets ($jj$). The first analysis targets
gluon--gluon fusion Higgs boson production and is sensitive to the CP properties of the $ggH$
effective coupling and the top-quark Yukawa coupling. The second analysis constrains Higgs
boson couplings to longitudinally and transversely polarised $W$ and $Z$ bosons in the VBF
production mode, assuming a CP-even Higgs boson. Both studies are based on proton–proton collision data
corresponding to an integrated luminosity of $36.1\;\mathrm{fb}^{-1}$ collected with the
ATLAS detector at $\sqrt{s} = 13~\TeV$ in the years 2015 and 2016. Representative leading-order
diagrams for the $\mathrm{ggF}$ + 2 jets and VBF production modes are depicted in
Figures~\ref{fig:FeynmandiagramsGGF}(a)--\ref{fig:FeynmandiagramsGGF}(c) and \ref{fig:FeynmandiagramsGGF}(d), respectively. \par
 
Both analyses use the shape of the signed azimuthal angle\footnote{ATLAS uses a right-handed coordinate system with its origin
at the nominal interaction point (IP) in the centre of the detector
and the $z$-axis along the beam pipe. The $x$-axis points from the
IP to the centre of the LHC ring, and the $y$-axis points
upwards. Cylindrical coordinates $(r,\phi)$ are used in the
transverse plane, $\phi$ being the azimuthal angle around the beam
pipe. The pseudorapidity is defined in terms of the polar angle
$\theta$ as $\eta = -\ln \tan(\theta/2)$. Angular distance is
measured in units of $\Delta R \equiv \sqrt{(\Delta\eta)^{2} +
(\Delta\phi)^{2}}$.} difference $\Delta\Phi_{jj}$ between the two leading hadronic jets
in the selected $H + 2$-jets candidate events to test for deviations from the SM expectations.
The angular difference is defined as $\Delta\Phi_{jj}=\phi_{j_1} - \phi_{j_2}$ if
$\eta_{j_1} > \eta_{j_2}$, and $\Delta\Phi_{jj}=\phi_{j_2} - \phi_{j_1}$ otherwise, where $j_1$ is the
highest-$p_\mathrm{T}$ jet and $j_2$ is the next-highest-$p_\mathrm{T}$ jet in the event.
The distribution of $\Delta\Phi_{jj}$ is probed in various disjoint kinematic regions, optimised for each analysis specifically.
 
The structure of this article is as follows. Section~\ref{sec:Theory} gives a short summary of
the theoretical frameworks used to study the CP properties of the Higgs boson's coupling to
top quarks and gluons, as well as its coupling to polarised electroweak bosons. The ATLAS
detector and the Monte Carlo and data samples used in these studies are discussed in
Section~\ref{sec:ATLASDet} and Section~\ref{sec:MCSimulation}, respectively. The
event selection and categorisation requirements are presented in Section~\ref{sec:ObjectAndEventSelection},
while the estimation of the various background processes is detailed in Section~\ref{sec:backEst}.
The theoretical and experimental uncertainties are presented in
Section~\ref{sec:Uncertainties}. Finally, the results
are discussed in Section~\ref{sec:Results}.
 
 
\section{Theoretical framework and methodology}\label{sec:Theory}
\begin{figure}[!ht]
\noindent \begin{centering}
\subfloat[$\mathrm{ggF}$ + 2 jets]{\includegraphics[width=0.5\textwidth]{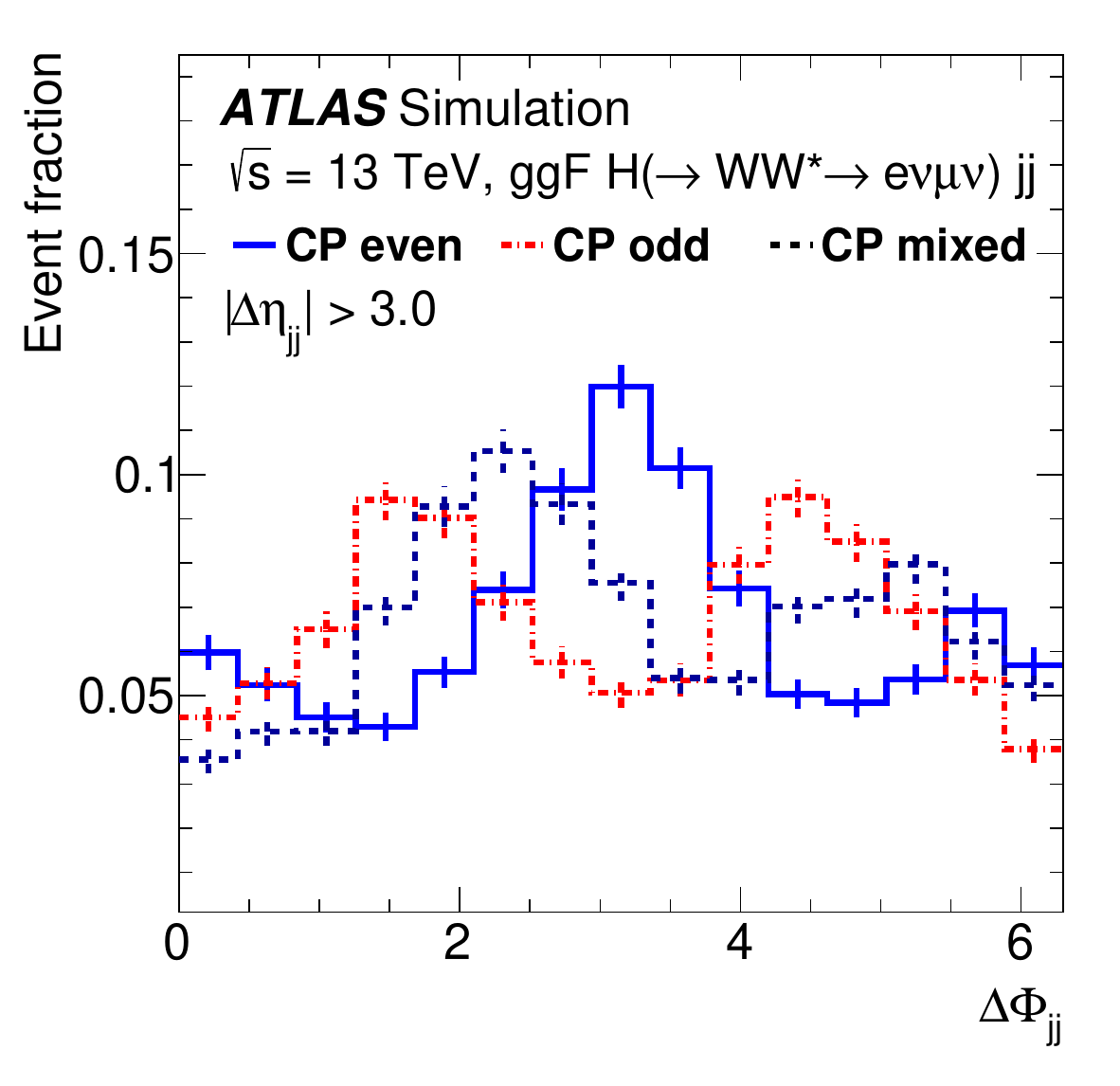}}
\subfloat[VBF]{\includegraphics[width=0.5\textwidth]{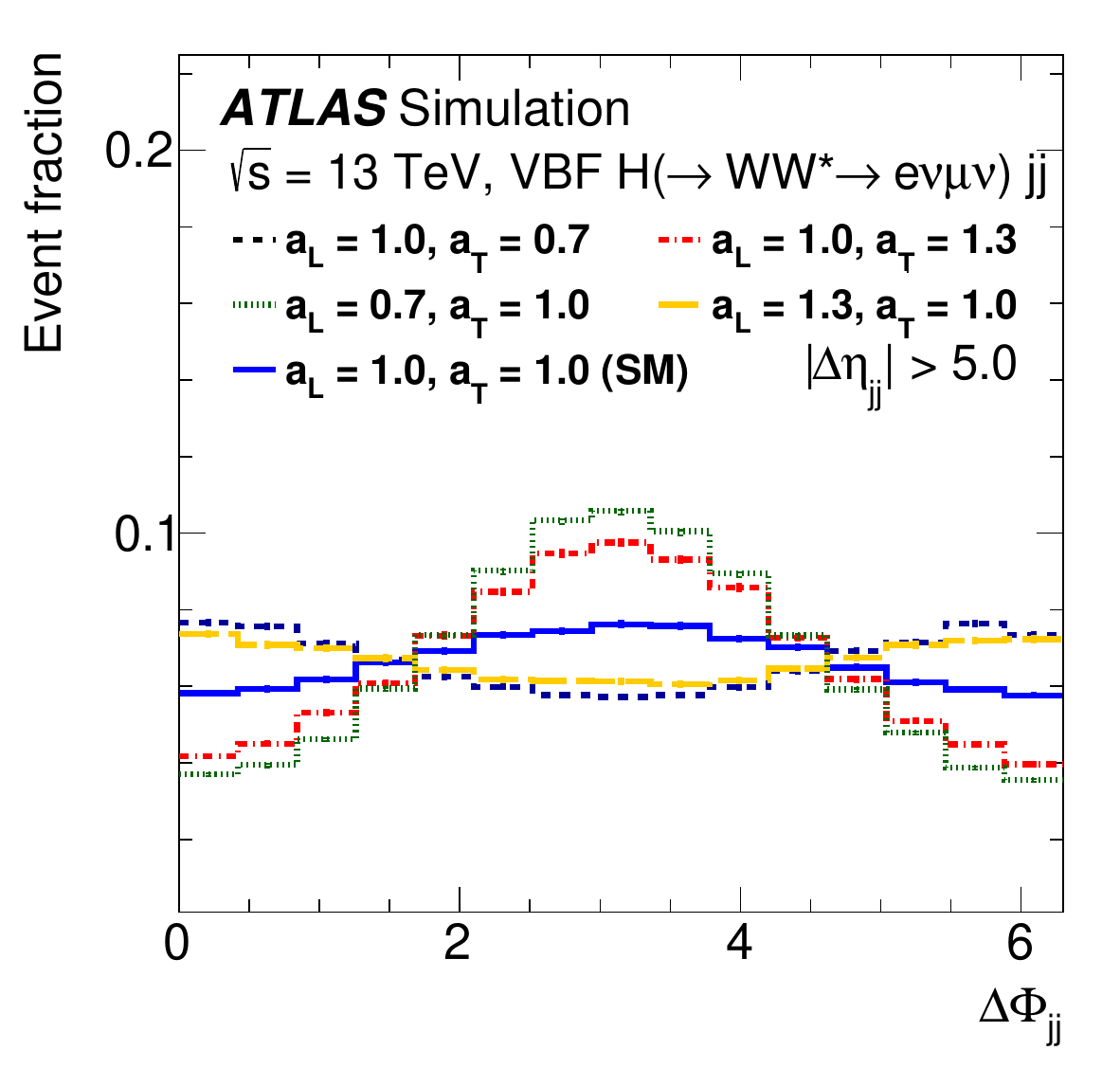}}
\par\end{centering}
\caption{Distributions of the signed $\Delta\Phi_{jj}$ observable shown for (a) CP-even, CP-odd and CP-mixed
benchmark models of the $\mathrm{ggF}$ + 2 jets production mode, and (b) various configurations of the
$a_\mathrm{L}$ and $a_\mathrm{T}$ parameters in VBF events. These comparisons are performed at the generator
level using the predictions of the \MGMCatNLOV{2.4.2} + \PYTHIAV{8.212} \cite{PYTHIA8} generators.}
\label{fig:DPhijjPlots}
\end{figure}

For the studies targeting beyond-the-SM (BSM) contributions to the top-quark Yukawa coupling and the effective
Higgs--gluon coupling, an effective field theory (EFT) framework is chosen to parameterise possible deviations from the SM.
The EFT operators probed in this article are provided by the Higgs Characterisation (HC) model~\cite{HiggsChar}, which is implemented in the
\MGMCatNLO generator \cite{MG5aMCatNLO,Frixione:2002ik}.
In the large top-quark-mass limit, $m_\mathrm{top}\rightarrow \infty$, the CP structure of the top-quark Yukawa coupling is
inherited by the effective Higgs--gluon interaction \cite{TopYukawaCP}. Thus, constraints on BSM contributions will be
directly set on the CP-even and CP-odd coupling strength modifiers of the effective Higgs--gluon interaction. The
effective Lagrangian that describes the Higgs--gluon interaction is expressed as
\begin{equation}\label{eq:LagrangianEFFHgg}
\mathcal{L}^\mathrm{loop}_{0} = -\frac{g^{}_{Hgg}}{4}\left(\kappa^{}_{gg} \cos(\alpha) G^a_{\mu\nu}G^{a,\mu\nu} + \kappa^{}_{gg} \sin(\alpha) G^a_{\mu\nu}\tilde{G}^{a,\mu\nu}\right) H \,,
\end{equation}
\noindent
 
where $G^a_{\mu\nu}$ is the gluon field strength tensor, $\tilde{G}^{a,\mu\nu}= G^a_{\rho\sigma}\,\varepsilon^{\mu\nu\rho\sigma}/2$
is the dual tensor, $g^{}_{Hgg}$ is the effective coupling for the SM CP-even $ggH$ interaction,
$\kappa^{}_{gg}$ is the coupling-strength scale factor for the effective Higgs--gluon interaction and $\alpha$ is the CP-mixing angle.
Interference between CP-even and CP-odd contributions affects the shape of the signed $\Delta\Phi_{jj}$ distribution,
but has no impact on the cross section of the $\mathrm{ggF}$ production mode, which is a function of $\kappa_{gg}^2\cos^2(\alpha)$
and $\kappa_{gg}^2 \sin^2(\alpha)$ only.
Three benchmark scenarios with different CP properties are defined in Table \ref{tab:benchmark}, and
the distribution of the signed $\Delta\Phi_{jj}$ observable is shown in Figure \ref{fig:DPhijjPlots}(a) for these
parameter choices.
 
\begin{table}[!ht]
\caption{Definition of the three benchmark scenarios used in the $\mathrm{ggF}$ + 2 jets analysis. The
parameter settings correspond to a CP-even (i.e.\ the SM hypothesis), a CP-odd, and a CP-mixed scenario.}
\centering
\small
\renewcommand{\arraystretch}{1.25}
\begin{tabular}{ll}
\hline
\hline
Scenario     & Parameters \\
\hline
CP-even (SM) & $\kappa^{}_{gg}=1$, $\cos(\alpha)=1$ \\
CP-odd       & $\kappa^{}_{gg}=1$, $\cos(\alpha)=0$ \\
CP-mixed     & $\kappa^{}_{gg}=1$, $\cos(\alpha)=\frac{1}{\sqrt{2}}$ \\
\hline
\hline
\end{tabular}
\label{tab:benchmark}
\end{table}
 
The analysis targeting $HVV$ couplings in Higgs boson production and decay
uses polarisation-dependent coupling-strength scale factors defined in Ref.~\cite{Brehmer:2014pka} as
\begin{equation}\label{eq:couplingsdefinition}
a_\mathrm{L} = \frac{g_{HV_\mathrm{L}V_\mathrm{L}}}{g_{HVV}}\,, \; a_\mathrm{T} =  \frac{g_{HV_\mathrm{T}V_\mathrm{T}}}{g_{HVV}}\, ,
\end{equation}
\noindent
where $g_{HVV}$ is the SM $HVV$ coupling strength and $g_{HV_\mathrm{L}V_\mathrm{L}}$ and
$g_{HV_\mathrm{T}V_\mathrm{T}}$ are the measured polarisation-dependent couplings.
 
The polarisations of the vector bosons in Eq.~\eqref{eq:couplingsdefinition} are defined in the
Higgs boson rest frame so that mixed-polarisation couplings $HV_\mathrm{L}V_\mathrm{T}$ do not contribute
to $\sigma_\mathrm{VBF} \cdot \mathcal{B}_{H\to WW}$. Other BSM effects are not considered.
Within the SM ($a_\mathrm{L}=a_\mathrm{T}=1$), the $HVV$ couplings are insensitive to the polarisations.
 
Since the polarisations depend on the measurement frame, the above description is not Lorentz
invariant and as such cannot be described in the Lagrangian framework.
Instead, the coupling strength modifiers $a_\mathrm{L}$ and $a_\mathrm{T}$ can be related to pseudo-observables (POs)~\cite{Greljo:2017spw}.
The POs considered in this article appear as $\kappa_{VV}$ and $\varepsilon_{VV}$ in the effective Lagrangian
\begin{equation}\nonumber
\mathcal{L} = \kappa_{VV} \left(\frac{2m_{W}^2}{v} H W^{+}_\mu W^{-\mu} + \frac{m_{Z}^2}{v} H Z_\mu Z^\mu \right) - \frac{\varepsilon_{VV}}{2v} \left(2HW^+_{\mu\nu} W^{-\mu\nu} + HZ_{\mu\nu} Z^{\mu\nu} + HA_{\mu\nu} A^{\mu\nu} \right) \,,
\end{equation}
\noindent
 
where in the SM $\kappa_{VV}=1$ and $\varepsilon_{VV}=0$. The universality of Higgs boson interactions with longitudinal $W$ and $Z$ bosons
follows from assuming custodial symmetry (see Eqs.~(33) and (35) in Ref.~\cite{Gonzalez_Alonso_2015}), no new physics in
the boson--fermion couplings $Wff$ and $Zff$, and a CP-even Higgs boson with CP-conserving interactions with vector bosons.
Since the study does not probe $HZ\gamma$ and $H\gamma\gamma$ interactions,
for simplicity a common coupling factor is assumed for Higgs-boson interactions with transversely polarised $W$ and $Z$ bosons, and photons.
 
The POs are related to the coupling-strength scale factors $a_\mathrm{L}$ and $a_\mathrm{T}$
via the following equations:
\begin{equation}\nonumber\label{eq:POs}
\kappa_{VV} = a_\mathrm{L} - \Delta_\mathrm{L}(q_1, q_2) \varepsilon_{VV}\,, \quad
\varepsilon_{VV} = \frac{a_\mathrm{T} - a_\mathrm{L}}{\Delta_\mathrm{T}(q_1, q_2)- \Delta_\mathrm{L}(q_1, q_2)} \,.
\end{equation}
\noindent
 
The functions $\Delta_\mathrm{L}(q_1, q_2)$ and $ \Delta_\mathrm{T}(q_1, q_2)$ depend on the momenta of electroweak bosons $q_{1}$ and $q_{2}$
(either in the production or in the decay) according to:
\begin{equation}\nonumber
\Delta_\mathrm{L} = \frac{m_H^2}{2 m_W^2} \frac{4q_1^2 q_2^2}{m_H^2(m_H^2 -q_1^2 -q_2^2)}\,, \quad
\Delta_\mathrm{T} = \frac{m_H^2}{2 m_W^2}\frac{m_H^2 -q_1^2 -q_2^2}{m_H^2}\,.
\end{equation}
\noindent
 
Based on the simulations of the VBF signal discussed in Section \ref{sec:MCSimulation}, $\Delta_\mathrm{L}(q_1, q_2)=0$ and $\Delta_\mathrm{T}(q_1, q_2)=2$ are found to be
good approximations, leading to the mapping
\begin{equation}\nonumber
\kappa_{VV}  \simeq a_\mathrm{L}\,, \quad \varepsilon_{VV} \simeq 0.5\cdot(a_\mathrm{T} - a_\mathrm{L})\,.
\end{equation}
\noindent
 
The above POs description focusses on VBF production and thus  differs from the one
used in Refs.~\cite{1708.02810, Aad:2020mkp}, in which couplings to leptons ($\varepsilon_{L}$, $\varepsilon_{R}$)
and to $Z$ bosons ($\kappa_{ZZ}$) are constrained using Higgs boson decays into
four leptons, inclusively with respect to the production channels. \par
 
This article consists of two studies. The first one places constraints on $\tan(\alpha)$
in the effective $Hgg$ coupling in $\mathrm{ggF}$ + 2 jets Higgs boson production, assuming standard $HVV$ couplings.
The second study constrains the $HVV$ parameters ($a_\text{L}$, $a_\text{T}$) and ($\kappa_{VV}$, $\varepsilon_{VV}$),
assuming a pure CP-even Higgs state with standard $Hgg$ coupling ($\kappa_{gg} \cos(\alpha) = 1$).
A non-SM $ggH$ coupling would negligibly impact the quantities determined in the VBF analysis and vice versa.
The constraints are derived from the rates of each production process as well as the distribution of signed
$\Delta\Phi_{jj}$, whose shape dependence on the coupling modifiers is shown in Figure~\ref{fig:DPhijjPlots}.
The distribution of $\Delta\Phi_{jj}$ is displayed in the full range $[0, 2\pi]$ to capture the asymmetry
for mixed CP interactions (shown in Figure~\ref{fig:DPhijjPlots}(a)) resulting from the interference between CP-even and CP-odd
contributions.\footnote{As the CP-even amplitude is positive under the transformation
$\left(\Delta\Phi_{jj}-\pi\right) \rightarrow - \left(\Delta\Phi_{jj}-\pi\right)$ and the CP-odd amplitude is negative under
this transformation, their interference is asymmetric whereas the individual squared amplitudes are positive.}
Figure~\ref{fig:DPhijjPlots}(b) shows the $\Delta\Phi_{jj}$ distribution for several choices of $a_\mathrm{L}$ and $a_\mathrm{T}$.
 
\section{ATLAS detector}\label{sec:ATLASDet}
The ATLAS detector~\cite{Aad:2008zzm} is a general-purpose particle
detector used to investigate a broad range of physics processes.
It includes an inner tracking detector (ID) surrounded by a
thin superconducting solenoid, electromagnetic and hadronic calorimeters,
and a muon spectrometer (MS) incorporating three large superconducting toroid magnets with eight coils each.
The ID consists of fine-granularity silicon pixel and microstrip
detectors, and a straw-tube tracker. It is immersed in a 2~T
axial magnetic field produced by the solenoid and provides precision tracking for charged
particles in the range $|\eta|<2.5$, where $\eta$ is the pseudorapidity of the particle.
The straw-tube detector also provides transition radiation measurements for electron
identification. The calorimeter system covers the pseudorapidity range
$|\eta| < 4.9$. It is composed of sampling calorimeters with either
liquid argon (LAr) or scintillator tiles as the active medium, and lead, steel, copper, or tungsten
as the absorber material. The MS provides muon identification and momentum measurements for
$|\eta| < 2.7$. The ATLAS detector has a two-level trigger system to
select events for further analysis.
 
 
\section{Datasets and Monte Carlo predictions}\label{sec:MCSimulation}
Candidate events in data are selected from the combined 2015 and 2016 ATLAS
$\sqrt{s}=13~\TeV$ $pp$ collision dataset in which all ATLAS
subdetectors were fully operational \cite{Aad:2019hmz}. The corresponding total integrated
luminosity \cite{LumiDetermination} is $36.1 \pm 0.8~\text{fb}^{-1}$. \par
 
The modelling of the gluon-induced production of Higgs bosons in association with jets was realised
using the \MGMCatNLOV{2.4.2} generator \cite{MG5aMCatNLO,Frixione:2002ik}, which provides a
calculation of the matrix element at next-to-leading-order (NLO) precision for $\mathrm{ggF}$ events with
up to two additional partons in the final state.
The calculations of the matrix element are based on the predictions of the HC
model, while the parton shower, hadronisation and underlying-event activity
were simulated with the \PYTHIA8.212 \cite{PYTHIA8} generator using
the A14 set of tuned parameters \cite{ATL-PHYS-PUB-2014-021}.
The cross-section calculation is based on the NNPDF3.0 \cite{Ball:2014uwa} NLO parton distribution function (PDF) sets.
In total, three different Monte Carlo samples were produced, corresponding to a CP-even, a CP-odd or a CP-mixed coupling
between Higgs bosons and gluons, following the recommendations from
Ref.~\cite{TopYukawaCP} and using the FeynRules model HC\_NLO\_X0\_UFO-heft \cite{Demartin:2015joa}.
The decay $H\rightarrow WW^{*}\rightarrow e\nu\mu\nu$ was modelled according to the SM. \par
 
For the measurement of the Higgs boson couplings to longitudinally and transversely polarised
$W$ and $Z$ bosons, the VBF production of the Higgs boson and its subsequent decays into
$W$ bosons were simulated at leading order (LO) in QCD using \MGMCatNLOV{2.4.2}.
The helicity amplitudes used in the matrix element generation of the Higgs boson
production and decay were modified to account for different Higgs coupling strengths in
the Higgs boson rest frame, following the prescription in Ref.~\cite{Brehmer:2014pka}.
Signal samples were produced for the following benchmark scenarios: $(a_\mathrm{L}, a_\mathrm{T}) \in \{$(1, 1), (1, 1.3), (1.3, 0.7), (1.3, 1), (0.7, 1.3)$\}$.
Parton shower effects were simulated with the \PYTHIA8.212 generator using the A14 set of tuned parameters.
The renormalisation and factorisation scales were both set to the sum of the transverse momenta of the jets, and the shower scale was set to 0.25
times the maximum $p_\mathrm{T}$ of the radiated partons. The SM (CP-even) sample described above was used for simulating the
gluon-fusion Higgs boson production in association with two jets.\par
 
For the studies of $\mathrm{ggF}$ + 2 jets production, the VBF production of the
Higgs boson is considered as a background and an additional sample was generated to model the SM prediction
for this process. In this case, \POWHEGBOXV{v2}~\cite{Alioli:2010xd} was used to model the matrix element at NLO in QCD
with the NNPDF3.0 NLO PDF set, while \PYTHIA8.212 was used to model parton shower effects. \par
 
The $q\bar{q}\rightarrow WH$ and $q\bar{q}\rightarrow ZH$ processes were generated with \POWHEGBOXV{v2} MiNLO \cite{POWHEG_MINLO}
interfaced to \PYTHIA8.186, with the AZNLO set of tuned parameters \cite{AZNLO_tune} for the underlying event, showering
and hadronisation. The $gg\rightarrow ZH$ process was simulated with \POWHEGBOXV{v2} + \PYTHIA8.186, with the
AZNLO set of tuned parameters for the underlying event, showering and hadronisation. For all $VH$ samples, the PDF4LHC15
PDF set \cite{PDF4LHC15} was used. The cross sections of the $VH$ production modes were fixed to SM predictions.
The $VH$ processes are considered as backgrounds to the studies of the $\mathrm{ggF}$ + 2 jets and VBF production modes. \par
 
Other production and decay modes of the Higgs boson were either fixed to SM predictions
(for $H\rightarrow\tau\tau$ decay) or neglected (for $t\bar{t}H$ and $b\bar{b}H$
associated production) due to their insignificant contributions to the phase-space
regions probed by the studies presented in this article. \par
 
The branching fractions for Higgs boson decays, calculated for a Higgs boson mass
$m_{H}=125\,\GeV$, were taken from the HDECAY program \cite{Djouadi:1997yw},
except for the signal samples used in the polarisation studies, for which the branching fractions
were taken from the predictions of the \MGMCatNLO generator\footnote{A common correction factor is applied to
all VBF signal samples to account for the difference between the branching fractions predicted by HDECAY
and those calculated by \MGMCatNLO for the SM scenario.}.
The inclusive $\mathrm{ggF}$ samples were normalised to match the state-of-the-art cross-section predictions
at next-to-next-to-next-to-leading-order (NNNLO) precision in QCD \cite{deFlorian:2016spz}.
The VBF samples were normalised to match cross-section predictions at next-to-next-to-leading-order (NNLO)
precision in QCD with NLO electroweak (EW) corrections, using the programs VBF@NNLO \cite{Cacciari:2015jma} and
HAWK \cite{Denner:2014cla}. \par
 
The main sources of background include events from the production of top quarks ($t\bar{t}$
and $Wt$), dibosons ($WW$, $WZ/WZ^{(*)}$, $ZZ$, $W\gamma^{(*)}$, $Z\gamma$) and single vector-bosons
($W$, $Z/\gamma^{*}$) plus jets. The production of a top-quark pair ($t\bar{t}$) was
modelled using the \POWHEGBOXV{v2} generator interfaced to \PYTHIA8.210 with the A14 set of
tuned parton shower parameters \cite{ATL-PHYS-PUB-2014-021}. The matrix elements were calculated at NLO precision in
QCD assuming a top-quark mass of $172.5~\GeV$. The NNPDF3.0 NLO PDF set was used and the $h_\mathrm{damp}$ parameter \cite{Jezo:2016ujg}
was set to $1.5$ times the top-quark mass. The $\ttbar$ production cross section was normalised to the predictions calculated with the \textsc{Top++2.0}
program to NNLO in perturbative QCD, including soft-gluon resummation
calculated to the next-to-next-to-leading logarithm (NNLL) \cite{Czakon:2011xs}.
The associated production of a single top quark and a $W$ boson ($Wt$) was generated with \POWHEGBOX interfaced to \PYTHIA6.428
for parton showering, using the \Perugia2012 tune~\cite{PerugiaTunes}. The matrix elements were calculated at NLO precision in QCD.
The cross-section calculation is based on the CT10 \cite{CT10PDF} PDF set. The diagram removal (DR)
scheme~\cite{Frixione:2008yi} was used to remove overlaps with the top-quark pair production process that occur at NLO in QCD. \par
 
The diboson production processes with $q\bar{q}$ and $qg$ initial states and leptonic final states were
simulated using \SHERPAV{2.2.2}~\cite{SHERPA,Bothmann_2019}. The cross-section calculation is based on the NNPDF3.0 NNLO PDF set and includes
all relevant off-shell components. Diagrams with up to one additional emission were
calculated with NLO precision in QCD, while diagrams with two or three parton emissions were
calculated with LO accuracy~\cite{ATL-PHYS-PUB-2017-005}. The various jet-multiplicity final
states were merged using the MEPS@NLO formalism \cite{Hoeche:2012yf} and a merging scale of
$Q_\mathrm{cut}=20~\GeV$. Loop-induced diboson processes that are initiated via the $gg$ production mode were
simulated at NLO in QCD using OpenLoops \cite{Cascioli:2011va} in \SHERPAV{2.1.1} and the CT10 NLO PDF set.
The production of dibosons with semileptonic decays, as well as the electroweak production of dibosons
in association with two jets, was modelled using \SHERPAV{2.1.1} and the CT10 PDF set. \par
 
The production of a $Z/\gamma^{*}$ boson in association with jets was modelled by \SHERPAV{2.2.1} with
the NNPDF3.0 NNLO PDF set. Diagrams with up to two additional parton emissions were simulated
with NLO precision in QCD, and those with three or four additional parton emissions to LO
accuracy. Matrix elements were merged with the \SHERPA parton shower using the MEPS@NLO
formalism with a merging scale of $Q_\mathrm{cut}=20~\GeV$ and the five-flavour numbering
scheme (5FNS). Contributions from the electroweak production of $q\bar{q}\rightarrow Z q\bar{q}$
events were considered, in which the matrix element was generated with up to one additional
emission beyond the first two partons using \SHERPAV{2.1.1} and the CT10 \cite{CT10PDF} PDF set.
Matrix elements were merged with the \SHERPA parton shower using the MEPS@NLO formalism
with the merging scale set to $Q_\mathrm{cut}=15~\GeV$. Contributions due to
processes containing a single $W$ boson produced in association with jets are estimated in a
purely data-driven approach. \par
The MC generators, PDF sets, and programs used to model the underlying event and parton shower
(UEPS) are summarised in Table~\ref{tab:mcsamples}.
The order of the perturbative prediction for each sample is also reported.\par
All simulated events were generated at a centre-of-mass energy of $\sqrt{s} = 13~\TeV$ and
then passed through the full ATLAS detector simulation \cite{fullsimATLAS,GEANT4}.
The simulated events were overlaid with additional inelastic $pp$ interactions that
were generated with \PYTHIA8.153 in order to match the pile-up conditions\footnote{An average of 13 (21)
interactions per bunch crossing were observed during the 2015 (2016) run.}
observed in the ATLAS data recorded during the 2015 and 2016 runs of the LHC.
 
\begin{table}[h]
\caption{Overview of the simulation tools used to generate signal and background processes, and to model the underlying event and parton shower.
The PDF sets are also summarised. The perturbative accuracy (in QCD and if relevant in EW corrections) of the total cross section is stated for each
process. The configuration marked with $^{(\star)}$ is used for the signal samples in the studies of the Higgs boson couplings to
longitudinally and transversely polarised $W$ and $Z$ bosons, while the configurations marked with $^{(\star\star)}$ are used as background
in the $\mathrm{ggF}$ + 2 jets analysis.
Alternative event generators and configurations used to estimate systematic uncertainties are shown in
parentheses. Further information about the alternative event generators and their configurations is given in Section \ref{sec:Uncertainties}, where MG5 is again used as an abbreviation for $\MADGRAPH5$.}
\centering
\small
\renewcommand{\arraystretch}{1.15}
\resizebox{1.0\textwidth}{!}{
\begin{tabular}{c c c c c}
\hline
\hline
Process            & Matrix element     & UEPS                & PDF set  &  Perturbative accuracy  \\
& \multicolumn{2}{c}{(alternative model)}  &          &  of total cross section \\
\hline
$\mathrm{ggF}$                                & MG5\_aMC@NLO 2.4.2    &  \PYTHIA8.212     &  NNPDF3.0 NLO   &  NNNLO QCD          \\
& \multicolumn{2}{c}{(MG5\_aMC@NLO 2.4.2 +  \HERWIG7.0.1)}    &           &         \\
VBF ${}^{(\star)}$                             & MG5\_aMC@NLO 2.4.2    &  \PYTHIA8.212     &  NNPDF3.0 NLO   &  NNLO QCD + NLO EW      \\
\hline
VBF ${}^{(\star\star)}$                         & \POWHEG-Box v2         & \PYTHIA8.212      &  PDF4LHC15 NLO  & NNLO QCD + NLO EW  \\
& \multicolumn{2}{c}{(MG5\_aMC@NLO 2.3.3  + \PYTHIA8.212)}   &                 &                \\
& \multicolumn{2}{c}{(\POWHEG-Box v2       + \HERWIG7.0.1)}   &                 &                \\
 
$VH$                                          & \POWHEG-Box v2     & \PYTHIA8.186     &  PDF4LHC15 NLO  & NNLO QCD + NLO EW  \\
\hline
$\ttbar$                                      & \POWHEG-Box v2     &  \PYTHIA8.210    &  NNPDF3.0 NLO   & NNLO+NNLL QCD\\
& \multicolumn{2}{c}{(\SHERPA 2.2.1)}  &                 &              \\
& \multicolumn{2}{c}{(\POWHEG-Box v2  + \HERWIG7.0.1)}     &              \\
$Wt$                                          & \POWHEG-Box v2     &   \PYTHIA6.428      &    CT10      & NLO QCD      \\
& \multicolumn{2}{c}{(MG5\_aMC@NLO 2.2.2   + \HERWIGpp)} &      &       \\
& \multicolumn{2}{c}{(\POWHEG-Box v2 + \HERWIGpp)} &      &              \\
$WZ/\gamma^{*}$, $ZZ/\gamma^{*}$      & \multicolumn{2}{c}{\SHERPA 2.2.2}  &  NNPDF3.0 NNLO  & NLO QCD \\
& \multicolumn{2}{c}{(MG5\_aMC@NLO 2.3.3 + \PYTHIA8.212)}  &      &         \\
$W\gamma, Z\gamma$                            & \multicolumn{2}{c}{\SHERPA 2.2.2}  &  NNPDF3.0 NNLO  & NLO QCD \\
& \multicolumn{2}{c}{(MG5\_aMC@NLO 2.3.3 + \PYTHIA8.212)}  &      &         \\
$qq, qg\rightarrow WW$                        & \multicolumn{2}{c}{\SHERPA 2.2.2}  &  NNPDF3.0 NNLO  & NLO QCD \\
& \multicolumn{2}{c}{(MG5\_aMC@NLO 2.3.3  + \PYTHIA8.212)}   &      &         \\
$gg\rightarrow WW$                            & \multicolumn{2}{c}{\SHERPA 2.1.1} &      CT10       & NLO  QCD \\
$Z/\gamma^{*}$                            & \multicolumn{2}{c}{\SHERPA 2.2.1}  &  NNPDF3.0 NNLO  & NNLO QCD \\
& \multicolumn{2}{c}{(MG5\_aMC@NLO 2.2.2 + \PYTHIA8.186)}  &      &      \\
\hline
\hline
\end{tabular}}
\label{tab:mcsamples}
\end{table}
 
 
\section{Event selection and categorisations}\label{sec:ObjectAndEventSelection}
Candidate events consistent with the final state $H\left(\rightarrow WW^{*}\rightarrow e\nu\mu\nu\right)$ + 2~jets
are selected. Events are retained for further analysis using single-lepton and dilepton triggers. The transverse momentum ($p_\mathrm{T}$)
thresholds range between $24\;\GeV$ and $26\;\GeV$ for single-electron triggers and between $20\;\GeV$ and $26\;\GeV$ for
single-muon triggers, depending on the run period \cite{ATLAS:2019dpa, ATLAS:2020gty, Aaboud:2016leb}, while the dilepton trigger
requires an electron with $p_\mathrm{T} > 17\;\GeV$ and a muon with $p_\mathrm{T} > 14\;\GeV$.
The combined efficiency of the single-lepton and dilepton triggers is $95\%$ in the fiducial
regions used in these analyses. \par
 
Candidate events are required to have at least one vertex having at least two associated tracks with
$p_\mathrm{T} > 400\;\MeV$. The vertex with the highest $\sum p^2_\mathrm{T}$ of the associated tracks
is taken as the primary vertex.
 
Electron candidates are reconstructed from tracks in the ID matched to
clusters \cite{Aaboud:2019ynx} of energy deposits in the electromagnetic (EM) calorimeter system. Electrons
are required to satisfy $|\eta| < 2.47$, excluding the transition region between calorimeters,
$1.37<|\eta|<1.52$. Muon candidates are reconstructed from combined tracks using information
from both the MS and the ID \cite{MuonIdentification}. This combination is based on an overall fit using the hits of
the ID track, the energy loss in the calorimeter, and the hits in the
muon spectrometer. The absolute pseudorapidity of the muon candidate is required to be lower than 2.5. \par
Various identification requirements including calorimeter- and track-based isolation criteria \cite{Aaboud:2019ynx,MuonIdentification}
are used to reduce the number of hadrons and soft leptons that arise from heavy-flavour decays and that are misidentified as prompt
leptons. The prompt-electron identification efficiencies range between $88\%$ and $94\%$ depending on the electron
$p_\mathrm{T}$ and $|\eta|$, while the prompt-muon identification efficiency is close to $95\%$ over the full
instrumented pseudorapidity range. \par
 
Jets are reconstructed from noise-suppressed topological clusters \cite{PERF-2014-07} of energy deposits in the calorimeter
system using the anti-$k_\mathrm{t}$ algorithm \cite{AntiKt,Fastjet} with a radius parameter of $R=0.4$. The jet
four-momentum is corrected using $p_\mathrm{T}$- and $\eta$-dependent scale
factors, which account for the calorimeter's non-compensating response, signal losses due to noise
threshold effects, energy lost in passive materials, and contributions from pile-up interactions \cite{JetMeasurement}. The absolute value of
the pseudorapidity of each jet is required to be lower than $4.5$ and the transverse momentum has to be at least $30\;\GeV$.
To reduce the contamination from jets originating from pile-up vertices, selection requirements on two independent
multivariate classifiers are applied to the selected jets. The first classifier is based on calorimeter and tracking
information and is applied to jets with $p_\mathrm{T} < 60\;\GeV$ and $|\eta| < 2.4$~\cite{ATLAS:2014cva}, while
the second classifier is based on jet shapes and topological jet correlations in pile-up interactions and is applied to
jets with $p_\mathrm{T}<50\;\GeV$ and $|\eta| >2.4$~\cite{PERF-2016-06}. \par
 
Jets containing $b$-hadrons are identified using the MV2c10 $b$-tagging algorithm~\cite{Aad:2015ydr,ATL-PHYS-PUB-2016-012}
with an operating point that has an overall efficiency of $85\%$, evaluated in simulated $t\bar{t}$
events. The corresponding overall rejection rate for jets originating from light-flavour
hadrons or gluons is 34, while the overall rejection rate for jets containing $c$-hadrons
is approximately 3. \par
 
The missing transverse momentum $\vec{E}^{\mathrm{miss}}_\mathrm{T}$ is defined as the negative vector sum of the $\vec{p}_\mathrm{T}$ of all the
selected leptons and jets, as well as all tracks compatible with the primary vertex but not associated with any of these
objects~\cite{ATLAS-CONF-2018-023}. \par
 
Ambiguities from overlapping reconstructed jet and lepton candidates are resolved as follows. If a reconstructed
muon shares an ID track with a reconstructed electron, the electron is removed. Reconstructed jets are
discarded if they are within a cone of size $\Delta R=0.2$ around an electron candidate or if they have fewer
than three associated tracks and are within a cone of size $\Delta R=0.2$ around a muon candidate. Electrons and muons
are removed if they are within $\Delta R=\min(0.4,0.04+10\;\GeV/p_\mathrm{T})$ of the axis of any surviving jet. \par
 
In order to be considered for the final analysis, candidate events must contain exactly two prompt leptons with
opposite electrical charge and different lepton flavour, with the higher-$p_\mathrm{T}$ (leading) lepton having $p_\mathrm{T}  > 22\;\GeV$
and the subleading lepton having $p_\mathrm{T} > 15\;\GeV$. The invariant mass of the dilepton system, $m_{\ell\ell}$,
must exceed $10\;\GeV$. At least one of the leptons must be matched to an object that triggered the recording of the event.
In the case that a dilepton trigger is solely responsible for the recording of an event, each lepton must be matched to one of the trigger-level objects.
In addition, events must contain at least two jets passing all selection requirements. To reduce backgrounds from top-quark production, events
are vetoed if they contain a $b$-tagged jet with a $\pt$ larger than $20\;\GeV$ ($N_{b\textrm{-jet}, p_\mathrm{T} > 20\;\GeV} = 0$).
The $Z(\rightarrow \tau\tau)$ + jets background is reduced by introducing an $m_{\tau\tau} < 66\;\GeV$
selection requirement. The observable $m_{\tau\tau}$ is calculated from the four-vectors of the two charged
leptons and the $\vec{E}^{\mathrm{miss}}_\mathrm{T}$ vector using the collinear approximation~\cite{Plehn:1999xi}. \par
 
Further selection requirements specific to the
topologies of the $\mathrm{ggF}$ + 2 jets and VBF signal processes are used to define the two signal regions (SRs).
In the $\mathrm{ggF}$ + 2 jets analysis, the angular distance between the two leading jets,
$\Delta R_{jj}$, is required to be larger than 1.0, the transverse momentum of the dilepton
system, $p_{\mathrm{T},\ell\ell}$, has to exceed $20\;\GeV$, and $m_{\ell\ell}$ has to
be below $90\;\GeV$. In addition, the transverse mass $m_\mathrm{{T}}$ of the Higgs boson candidate
must be below $150\;\GeV$. This transverse mass is defined as
$m_\mathrm{{T}} = \sqrt{ {(E_{\ell\ell} + E_\mathrm{T}^\mathrm{miss})}^2 - {|\vec{p}_{\mathrm{T},\ell\ell} + \vec{E}_\mathrm{{T}}^\mathrm{miss}|}^2 }$ with
$E_{\ell\ell} = \sqrt{|\vec{p}_{\mathrm{T},\ell\ell}|^2 + m_{\ell\ell}^2 }$ and $\vec{p}_{\mathrm{T},\ell\ell}$ the combined dilepton
momentum vector in the transverse plane. The selection requirement placed on $p_{\mathrm{T},\ell\ell}$ reduces contributions
from the $Z$ + jets background, while the requirements on $m_{\ell\ell}$ and $m_\mathrm{{T}}$ decrease the top-quark background.
In the VBF analysis, an `outside lepton veto' is applied, which requires the two leptons to be within the rapidity gap spanned
by the two leading jets, and a `central jet veto' rejects events with additional jets with $p_\mathrm{T} > 20\;\GeV$ in
the rapidity gap between the two leading jets. Table~\ref{tab:SignalRegionsDef} summarises the selection requirements used to define
the SRs of the $\mathrm{ggF}$ + 2 jets and VBF analyses. \par
 
Boosted decision trees (BDTs) are used in both analyses to further distinguish between the signal processes
and the most dominant background processes. In the $\mathrm{ggF}$ + 2 jets analysis, the discriminating observables used by the BDT
are $m_{\ell\ell}$, $m_\mathrm{T}$, $p_{\mathrm{T},\ell\ell}$, and the azimuthal angle $\Delta\phi_{\ell\ell}$ between the two leptons,
as well as the observables $\min \Delta R (\ell_{1},j_{i})$ and $\min \Delta R (\ell_{2},j_{i})$, defined as the minimum distance between the
leading or subleading lepton and the two tagging jets. The $m_{\ell\ell}$ and $m_\mathrm{T}$ observables provide the best separation between
the signal and background processes, and are therefore the most important inputs to the BDT. The sample used to train the BDT consists of
the sum of simulated $\mathrm{ggF}$ + 2 jets events stemming from the three benchmarks models defined in Table \ref{tab:benchmark} as signal, and the
sum of the top quark, diboson and $Z$ + jets processes as background. Neither the input observables nor the BDT
response show any significant separation between the three CP benchmark scenarios.
In the VBF analysis, the same BDT is used as described in Ref.~\cite{atlas-higg-2016-07}, where $m_{\ell\ell}$, $\Delta\phi_{\ell\ell}$, and $m_\mathrm{{T}}$
are used in addition to the dijet invariant mass $m_{jj}$, the rapidity difference $\Delta y_{jj}$ between the two leading jets, the lepton
centrality ($\sum_{\ell} C_\ell$ where $C_\ell = |2\eta_\ell - \sum_j \eta_j|/\Delta\eta_{jj}$ quantifies the position of a lepton
relative to the two leading jets), the sum of the invariant masses of
all four possible lepton--jet pairs, $\sum_{\ell,j} m_{\ell,j}$, and the total transverse momentum $p^\text{tot}_\mathrm{T}$, defined as the
magnitude of the vectorial sum of all selected objects. The most discriminating observables used by the BDT of the VBF analysis are
$m_{jj}$ and $\Delta y_{jj}$.
 
\begin{table}[!h]
\caption{Event selection criteria used to define the signal regions for the $\mathrm{ggF}$ + 2 jets and VBF event categories.}
\centering
\small
\renewcommand{\arraystretch}{1.2}
\begin{tabular}{c | c | c}
\hline
\hline
&        $\mathrm{ggF}$ + 2 jets                                 &       VBF                                          \\
\hline
 
&  \multicolumn{2}{c}{Two isolated, different-flavour leptons ($\ell = e, \mu$) with opposite charge}          \\
\multirow{2}{*}{Preselection} &  \multicolumn{2}{c}{$p^\mathrm{lead}_\mathrm{T}>22\;\GeV$, $p^\mathrm{sublead}_\mathrm{T}>15\;\GeV$}                \\
&  \multicolumn{2}{c}{$m_{\ell\ell} > 10\;\GeV$}                                                                \\
&  \multicolumn{2}{c}{$N_{\mathrm{jet}}\geq 2 $}                                                               \\
\hline
&  \multicolumn{2}{c}{$N_{b\textrm{-jet},p_\mathrm{T} > 20\;\GeV} = 0$}                                     \\
&  \multicolumn{2}{c}{$m_{\tau\tau} < 66\;\GeV$}                                                             \\
\multirow{2}{*}{Background rejection} & \hspace{1.6cm} $\Delta R_{jj} > 1.0$   \hspace{1.6cm}    &                                                         \\
& $p_{\mathrm{T},\ell\ell}> 20\;\GeV$                     &        central jet veto                                 \\
& $m_{\ell\ell} < 90\;\GeV$                               &        outside lepton veto                              \\
& $m_\mathrm{T} < 150\;\GeV$                              &                                                         \\
\hline
\multirow{2}{*}{BDT input variables}  &  $m_{\ell\ell}$, $m_\mathrm{T}$, $p_{\mathrm{T},\ell\ell}$, $\Delta\phi_{\ell\ell}$   &     $m_{jj}$, $\Delta y_{jj}$, $m_{\ell\ell}$, $m_\mathrm{T}$, $\Delta\phi_{\ell\ell}$  \\
&  $\min \Delta R (\ell_{1},j_{i})$, $\min \Delta R (\ell_{2},j_{i})$                    &     $\sum_{\ell} C_{\ell}$, $\sum_{\ell,j} m_{\ell,j}$, $p^\text{tot}_\mathrm{T}$    \\
\hline
\hline
\end{tabular}
\label{tab:SignalRegionsDef}
\end{table}
 
\section{Background estimation}\label{sec:backEst}
The background contamination in the SRs originates from various processes including the production of top-quark pairs
($t\bar{t}$), single top quarks ($Wt$), non-resonant dibosons ($WW$, $WZ/\gamma^{*}$, $ZZ/\gamma^{*}$, $W\gamma$, or $Z\gamma$), and
Drell--Yan $Z/\gamma^{*}$ (primarily in the decay $Z\rightarrow\tau\tau$).
Other background contributions arise from $W$ + jets and multijet production
with misidentified leptons, which originate either from decays of heavy-flavour hadrons or
from jets mimicking prompt-lepton signatures in the detector. A small but non-negligible contribution to
the signal region arises from $VH$ production processes and events including $H\rightarrow\tau\tau$ decays.
These two sources are summarised as one component referred to as ``Other Higgs'', which
is treated as a background to the studies of the $\mathrm{ggF}$ + 2 jets and VBF production modes. \par
Dedicated control regions (CRs), which do not overlap with the signal region, are used to constrain
the normalisation of the most dominant background processes. A top CR is used to
correct the normalisation of the combined $t\bar{t}$ and $Wt$ backgrounds.
A CR is employed to adjust the normalisation of the $Z(\to\tau\tau)$ + 2 jets production.
In the $\mathrm{ggF}$ + 2 jets analysis, an additional $WW$ CR is used.
The selection criteria used to define the various CRs are detailed in Table~\ref{tab:ControlRegionsDef}.
 
\begin{table}[!h]
\caption{Event selection criteria used to define the various control regions for the
$\mathrm{ggF}$ + 2 jets and VBF event categories. Only the changes relative to the
signal region definitions (see Table~\ref{tab:SignalRegionsDef}) are stated.}
\centering
\small
\renewcommand{\arraystretch}{1.2}
\begin{tabular}{c | c | c}
\hline
\hline
Control region                              &   $\mathrm{ggF}$ + 2 jets                                     &       VBF              \\
\hline
Top CR                                       &   $N_{b\textrm{-jet},(p_\mathrm{T} > 30\;\GeV)} = 1$  &  $N_{b\textrm{-jet},(p_\mathrm{T} > 20\;\GeV)} = 1$  \\
\hline
\multirow{2}{*}{$Z\rightarrow \tau\tau$ CR}  &   \multicolumn{2}{c}{$| m_{\tau\tau} - m_{Z} | \leq 25\;\GeV$}                 \\
&   $p_{\mathrm{T},\ell\ell}$ requirement is omitted        &    $m_{\ell\ell}$ < 80\;\GeV \\
\hline
\multirow{2}{*}{$WW$ CR}                     &   $m_{\ell\ell} > 90\;\GeV$                           &  \multirow{2}{*}{----} \\
&   $m_\mathrm{T}$ requirement is omitted                   &                        \\
\hline
\hline
\end{tabular}
\label{tab:ControlRegionsDef}
\end{table}

Diboson backgrounds other than $WW$ are estimated using MC simulations, while the contributions
from background processes containing misidentified leptons are estimated with a data-driven
technique \cite{atlas-higg-2016-07}. For this purpose, a control sample is defined using events with one identified
lepton and one lepton failing the nominal object definition requirements but passing looser
requirements (referred to as ``anti-identified''). The contribution of the misidentified-lepton
background to the signal region is estimated by scaling the control sample via
$p_\mathrm{T}$- and $\eta$-dependent extrapolation factors, which are defined as the ratio of
the identified leptons to anti-identified leptons and are determined using $Z$ + jets and multijet
data samples.  \par
 
The composition of the signal and control regions is detailed in Table \ref{tab:MuFit2D_EventYieldsPostFit_unblinded} for
the $\mathrm{ggF}$ + 2 jets analysis and in Table \ref{tab:VBF_ALfit_EventYieldsPostFit} for the VBF analysis.
For both analyses, the largest background contribution to the signal region stems from top-quark pair production
with around $50\%$ of all selected events. The next leading contributions to the signal regions are from $WW$ and $Z$ + jets
production, which contribute each with roughly $20\%$ of all selected events. \par
 
The procedure described in Section \ref{sec:Results} is used to obtain the normalisation factors (NFs)
of the dominant background processes from a combined SR + CR fit to data, where each background
normalisation is correlated across all regions. In the $\mathrm{ggF}$ + 2 jets analysis, the
normalisations of the top quark, $Z$ + jets and $WW$ backgrounds are determined simultaneously.
The $WW$ CR has approximately equal contributions from $WW$ and top processes, so there is a moderate
anti-correlation between the NFs of the $WW$ and top-quark backgrounds. The $WW$ CR nonetheless
constrains the sizeable $WW$ modelling uncertainty. In the VBF analysis, the normalisations of the top quark
and $Z$ + jets backgrounds are determined from a simultaneous fit to their dedicated CRs. The normalisation factors
obtained from the combined SR + CR fit to data are summarised in Table~\ref{tab:PostNormalisationFactors}
separately for the $\mathrm{ggF}$ + 2 jets and VBF analyses.
 
\begin{table}[!h]
\caption{Post-fit NFs and their uncertainties for the $Z$ + jets, top and $WW$ backgrounds.
Both sets of normalisation factors differ slightly depending on which (B)SM model is tested, but are consistent within their total uncertainties.}
\centering
\small
\renewcommand{\arraystretch}{1.6}
\begin{tabular}{c c c c}
\hline
\hline
Phase space             &   NF$^{Z\mathrm{ + jets}}$       & NF$^\mathrm{top}$          &  NF$^{WW}$              \\
\hline
$\mathrm{ggF}$ + 2 jets  &   $0.85^{+0.09}_{-0.09}$     & $1.05^{+0.06}_{-0.05}$  &  $1.0^{+0.7}_{-0.4}$       \\
VBF                      &   $0.94^{+0.21}_{-0.18}$     & $1.00^{+0.06}_{-0.05}$  &     ----                   \\
\hline
\hline
\end{tabular}
\label{tab:PostNormalisationFactors}
\end{table}
 
\section{Systematic uncertainties}\label{sec:Uncertainties}
The effects of the systematic uncertainties on the expected signal and background yields in the various signal and control
regions are evaluated following the procedures in Ref.~\cite{atlas-higg-2016-07}. In addition, the effects of the uncertainties
on the shapes of the $\Delta\Phi_{jj}$ and BDT response distributions are considered. They are evaluated by individually
comparing the nominal distribution with those corresponding to up and down variations of a particular uncertainty. \par
 
The sources of uncertainty are grouped into two categories: experimental and theoretical. The dominant
experimental uncertainties for both analyses are related to the $b$-tagging efficiency \cite{Aaboud:2018xwy},
the jet energy scale and resolution \cite{ATLAS:2017bje}, the modelling of pile-up activity, and the
estimation of the misidentified-lepton background \cite{atlas-higg-2016-07}.
Smaller uncertainties are due to the lepton momentum scale and resolution, the lepton identification and
isolation criteria \cite{Aaboud:2019ynx,Aaboud:2018ugz,MuonIdentification}, the measurement of missing
transverse momentum \cite{ATLAS-CONF-2018-023}, and the luminosity measurement \cite{ATLAS-CONF-2019-021}. The
luminosity uncertainty is only applied to those processes that are normalised to theoretical
predictions. \par
 
Theoretical uncertainties are assessed by comparing nominal and alternative event generators and
UEPS models, as indicated in Table \ref{tab:mcsamples}. In general, the perturbative precision and the PDF sets used in
these alternative configurations match those of the nominal generators (unless explicitely stated).
Uncertainties due to the PDF set parameterisation are evaluated using replica sets, and uncertainties
due to missing higher orders are evaluated by varying appropriate scale parameters, as described below. \par
 
The uncertainties in modelling the $t\bar{t}$ background
are derived as follows. To assess potential differences in the matching between the matrix element and parton
shower, the predictions of the nominal generator set-up are compared with those of the \SHERPAV{2.2.1} generator.
The cross-section calculation of this alternative generator set-up is based on the NNPDF3.0 NNLO PDF set and
includes diagrams with up to one additional emission at NLO precision in QCD, while diagrams with two, three or four
parton emissions are described at LO accuracy. Parton shower modelling uncertainties are derived by
replacing \PYTHIA8.210 by \HERWIG~7.0.1 \cite{HERWIG7} (which uses the H7-UE-MMHT set of tuned parameters \cite{H7-UE-MMHT-tune})
and comparing the corresponding yields and shapes with those from the nominal set-up. The uncertainty due to neglected higher orders in QCD
is estimated by simultaneously increasing (decreasing) the renormalisation and factorisation scales $\muR$ and
$\muF$ by a factor of 2 (0.5), and by setting the $h_\mathrm{damp}$ parameter to 1.0 (3.0) times the top-quark mass.
For $Wt$ production, uncertainties in the matching between the matrix element and parton shower are evaluated by comparing
the predictions of \POWHEGBOXV{v2} + \HERWIGpp \cite{HERWIGpp} with those of MG5\_aMC@NLO~2.2.2 + \HERWIGpp, while the parton
shower model uncertainties are estimated by comparing \POWHEGBOXV{v2} + \PYTHIA6.428 with \POWHEGBOXV{v2} + \HERWIGpp.
In these samples, the UE-EE-5-CTEQ6L1 set of tuned parameters \cite{HERWIGpp} is used to configure the \HERWIGpp generator.
In addition, the nominal configuration for the $Wt$ process is compared with an alternative approach in which the diagram
subtraction scheme \cite{White:2009yt} is used instead of the DR scheme. The uncertainty due to neglected higher orders in QCD is
estimated by increasing (decreasing) the renormalisation and factorisation scales by a factor of 2 (0.5) relative to their nominal value. \par
 
The uncertainties in modelling the production of dibosons with jets are evaluated by comparing the predictions of the
\SHERPAV{2.2.2} and MG5\_aMC@NLO~2.3.3  + \PYTHIA8.212 generators, where the latter provides NLO precision in QCD for the
simulation of production modes with up to one parton in addition to the diboson system \cite{ATL-PHYS-PUB-2017-005}.
The cross-section calculation is based on the NNPDF3.0 PDF set, and the A14 set of tuned parameters is used for the simulation of
the parton shower. In this generator set-up, the FxFx merging scheme \cite{FxFxmerging} is used to remove
overlaps between the partonic configuration produced during the simulation of the matrix element and the
parton shower, using a merging scale of $20\;\GeV$. For the predictions of the $WZ/\gamma^{*}$, $ZZ/\gamma^{*}$,
and $WW$ processes, variations of the matching scale are also considered, where the nominal value, $20\;\GeV$, is changed to $30\;\GeV$ and $15\;\GeV$.
In addition, the effects of QCD factorisation and renormalisation scale variations are considered by
individually varying $\muR$ and $\muF$ by a factor of 2 or 0.5. Six combinations are
considered: $\left(\muR, \muF\right)$ = $\left(0.5, 0.5\right)$, $\left(0.5, 1.0\right)$, $\left(1.0, 0.5\right)$,
$\left(1.0, 2.0\right)$, $\left(2.0, 1.0\right)$, and $\left(2.0, 2.0\right)$ times their nominal value. Finally, the uncertainty due to the QCD scales is obtained as
the largest effect of these six variations of \muR and \muF from their nominal value. \par
 
For the $Z$ + jets background, matrix element and parton shower
uncertainties are both accounted for by comparing the predictions of the \SHERPAV{2.2.1} generator with those of
MG5\_aMC@NLO~2.2.2 + \PYTHIA8.186, which provides matrix elements at LO accuracy with up to four final-state partons.
The alternative event generator set-up is based on the NNPDF2.3 LO PDF set for the cross-section calculations and the A14 set
of tuned parameters for the simulation of the parton shower. The effect of QCD scale variations on the $Z$ + jets background are
determined in the same way as for the diboson plus jets backgrounds. \par
 
The uncertainties in modelling the ggF and VBF production modes of the Higgs boson are evaluated as follows.
Uncertainties due to the parton shower model for the $\mathrm{ggF}$ + 2 jets process are determined by comparing the predictions of the
nominal generator configuration with those of the MG5\_aMC@NLO~2.4.2 + \HERWIG~7.0.1 generators, where the
H7-UE-MMHT set of tuned parameters is used for the simulation of the parton shower.
The effects of parton shower model uncertainties on the VBF signal are calculated by varying
the shower scale up and down by a factor of 2 in \PYTHIA8.
The effects of QCD scale variations are determined for
the $\mathrm{ggF}$ + 2 jets and VBF processes in the same way as for the vector boson
plus jets backgrounds. Uncertainties in the ggF and VBF production cross sections and jet bin migration effects for the $\mathrm{ggF}$
process are accounted for following the recommendations in Ref.~\cite{deFlorian:2016spz}.
In the $\mathrm{ggF}$ + 2 jets analysis, further uncertainties are considered for the VBF background. Uncertainties due to the
matching of the matrix element and the parton shower are evaluated by comparing the predictions of \POWHEGBOXV{v2} + \PYTHIA8.212 with
the predictions of MG5\_aMC@NLO~2.3.3  + \PYTHIA8.212, while uncertainties due to the parton shower model are derived by comparing the predictions
of \POWHEGBOXV{v2} + \PYTHIA8.212 with those of \POWHEGBOXV{v2} + \HERWIG~7.0.1, where the H7-UE-MMHT set of tuned parameters
is used for the simulation of the parton shower. In the VBF study, an additional shape uncertainty is applied to the
$\Delta\Phi_{jj}$ and BDT response distributions of all backgrounds to account for a non-closure between data and the simulations
in the dijet invariant mass spectrum. 
This non-closure is on the order of a few percent at low $m_{jj}$ values and ranges up to $15\%$ for $m_{jj}$ values above $1\,\TeV$. Its impact on the
final results is small with respect to the statistical uncertainties. In the VBF analysis, parton shower model uncertainties on the VBF signal are
calculated by setting the \PYTHIA shower scale factor to $2$ or $1/2$. \par
 
Uncertainties due to the modelling of PDFs are evaluated for the signal processes and all relevant backgrounds by comparing the predictions
of the nominal PDF set with those of the CT14 and MMHT2014 PDF sets and then comparing the maximum difference with the difference from
the root-mean-square spread of the NNPDF3.0 replica sets. The larger of the two is taken as the uncertainty. The uncertainties in
the signal processes are evaluated for the SM hypotheses and are extrapolated to the various BSM scenarios while assuming that
QCD scale, PDF and parton shower model effects factorise with the modulations of the $\Delta\Phi_{jj}$ shape and cross sections due
to BSM contributions. These modelling uncertainties are treated as fully correlated between the SM and BSM hypotheses. \par
 
The most significant theoretical uncertainties are related to the modelling of the top-quark and $WW$ backgrounds and the $\mathrm{ggF}$
process. Tables~\ref{tab:TanAlphaFit_ShapeRateOnData} and ~\ref{tab:PO_ShapeRateOnData} rank the various uncertainties and show
their impact on the studies of the $\mathrm{ggF}$ + 2 jets and VBF processes. Both studies are dominated by statistical
uncertainties. Some systematic uncertainties have a significant impact on the determination of
the post-fit normalisation factors (presented in Table~\ref{tab:PostNormalisationFactors}). In the $\mathrm{ggF}$ + 2 jets
analysis, there are significant anti-correlations between the normalisation of the top-quark background and the uncertainty in
the $b$-tagging efficiency. The normalisation factor of the $WW$ background is significantly anti-correlated with both
the uncertainty in the $b$-tagging efficiency and the effect of the QCD scale uncertainty on the $WW$ backgrounds, while the normalisation
factor of the $Z$ + jets background is significantly anti-correlated with the jet energy scale uncertainties. In the VBF analysis,
no significant anti-correlations are observed between the normalisation factors and the various uncertainties.
 
\section{Results}\label{sec:Results}
 
A maximum likelihood (ML) fit is used for the statistical interpretation of the results from
both the $\mathrm{ggF}$ + 2 jets and VBF analyses.
Fits are simultaneously performed on all considered SRs and CRs of each analysis independently
in order to constrain the normalisation of backgrounds and the nuisance parameters (NPs) describing
the systematic uncertainties. Each systematic variation enters the fit as an individual nuisance parameter.
Correlations between systematic uncertainties are maintained across
processes and channels. Due to the presence of bins with low event yields, all the NPs describing
the systematic uncertainties are incorporated into the fit using a log-normal constraint, such that
expected event counts remain positive for all values of the corresponding NPs.
Asimov datasets \cite{Cowan:2010js} are used to study the expected performance of each fit. \par
 
Parameter morphing \cite{morph, Interpolation} is used to interpolate and extrapolate from a
small set of $\kappa^{}_{gg}\cos(\alpha)$ and $\kappa^{}_{gg}\sin(\alpha)$ (or $a_\mathrm{L}$ and $a_\mathrm{T}$) coupling benchmarks to a
large variety of coupling scenarios. The input distributions to the morphing are normalised
to their expected cross sections. \par
 
The final results are obtained by applying the maximum-likelihood procedure individually to each
coupling parameter hypothesis, where the background prediction is only affected by changes to
nuisance parameters in the minimisation. A negative log-likelihood
(NLL) curve is constructed as a function of the relevant coupling parameters. The best estimate
for the parameter of interest is obtained at the point where the NLL curve reaches its minimum.
In addition, central confidence intervals are determined from the appropriate deviation of the
NLL from its minimum. \par
 
\subsection{$\mathrm{ggF}$ + 2 jets category}
 
The ML fits that constrain BSM effects in the effective Higgs--gluon coupling
use the distribution of the signed $\Delta\Phi_{jj}$ observable as input, divided into 12 categories.
These 12 different categories are obtained by splitting the signal region into three BDT score
intervals times four $|\Delta\eta_{jj}|$ intervals.\footnote{The split into BDT score regions aims to maximise
the signal to background ratio, while the split into $|\Delta\eta_{jj}|$ regions is motivated
by the fact that the separation between the various CP hypotheses for the signed
$\Delta\Phi_{jj}$ distribution increases with higher $|\Delta\eta_{jj}|$ values.}
The bin boundaries for the BDT score and $|\Delta\eta_{jj}|$ intervals that define the
categories are $[0.1,0.4,0.7,1.0]$ and $[0.0,1.0,2.0,3.0,\infty]$,
respectively. Finally, the event yields from the top, $Z\rightarrow\tau\tau$ and
$WW$ CRs, as well as the event yields within the low-BDT-score intervals
$w_\mathrm{BDT}\in[-1.0,-0.3]$ and $w_\mathrm{BDT}\in[-0.3,0.1]$ in each $|\Delta\eta_{jj}|$ region
are included in the ML fit. These regions provide constraints on the normalisations of the
top quark, $Z$ + jets and $WW$ + jets backgrounds, which are free to float in the fit, and
help to reduce the impact of the $b$-tagging uncertainties. All other background contributions are set to
their respective SM predictions, but are allowed to vary within their uncertainties. \par
For the measurement of the signal strength parameter for $\mathrm{ggF}$ + 2 jets events, $\mu^{\mathrm{ggF + 2 jets}}$,
the ML fit uses the same five BDT-score and four $|\Delta\eta_{jj}|$ intervals as those used for the studies of the
effective Higgs--gluon coupling. In addition to these intervals, the event yields from the top, $Z\rightarrow\tau\tau$, and
$WW$ CRs are included in the fit. The normalisations of the top quark, $Z$ + jets,
and $WW$ + jets backgrounds are free to float in the fit, while all
other background contributions are set to their respective SM predictions, but are
allowed to vary within their uncertainties. \par
 
Four different fits are performed in the $\mathrm{ggF}$ + 2 jets event category: \par
 
\begin{itemize}
\item{ The signal strength parameter $\mu^\mathrm{ggF + 2 jets}$ for the $\mathrm{ggF}$ + 2 jets
signal process is determined. This parameter is defined as the ratio of the measured signal yield to
that predicted by the SM.}
\item In order to constrain BSM effects in the effective Higgs--gluon coupling,
$\tan(\alpha)$ is scanned. Two different configurations
are used in the ML procedure:
\begin{itemize}
\item{The normalisation of the signal process is unconstrained such that the analysis only
exploits the shape information of the fit input distribution to distinguish between the different
CP scenarios.}
\item{The signal normalisation is constrained to the model predictions. Thus both the shape and rate
information are considered for each CP scenario.}
\end{itemize}
Using rate information in addition to the shape in the fit increases the sensitivity
to distinguish between the various benchmark points\footnote{BSM effects can have significant contributions to the production rates of the signal process. For example, the production cross section of the CP-odd benchmark scenario defined in Table 1 is enhanced by a factor of around 2.3 with respect to the SM expectations,  while the CP-mixed benchmark scenario has a cross section of around 1.6 times the SM expectation.}. However, the rate can be
affected by multiple BSM effects, while the shape isolates CP-dependent variations.
\item{A simultaneous fit of the coupling-strength scale factors $\kappa^{}_{gg}\cos(\alpha)$ and $\kappa^{}_{gg}\sin(\alpha)$
is performed. This study exploits both shape and rate information.}
 
\end{itemize} \par
 
The inclusive $\mathrm{ggF}$ production process is split into two components, one describing contributions from $\mathrm{ggF}$ + 0/1 jets
and the other $\mathrm{ggF}$ + 2 or more jets, where only the latter one is sensitive to the CP nature of the effective Higgs--gluon coupling.
The treatment of these two components varies between the different fit types. For scans over $\tan(\alpha)$ where only the shape is used,
the normalisation of the $\mathrm{ggF}$ + 2 jets component floats freely in the fit, while the normalisation of the
$\mathrm{ggF}$ + 0/1 jets component may vary taking into account its uncertainties. The normalisations of both components are scaled
consistently to the model predictions when rates are used in the fit.
 
\begin{figure}[!h]
\noindent \begin{centering}
\subfloat{\includegraphics[width=0.65\textwidth]{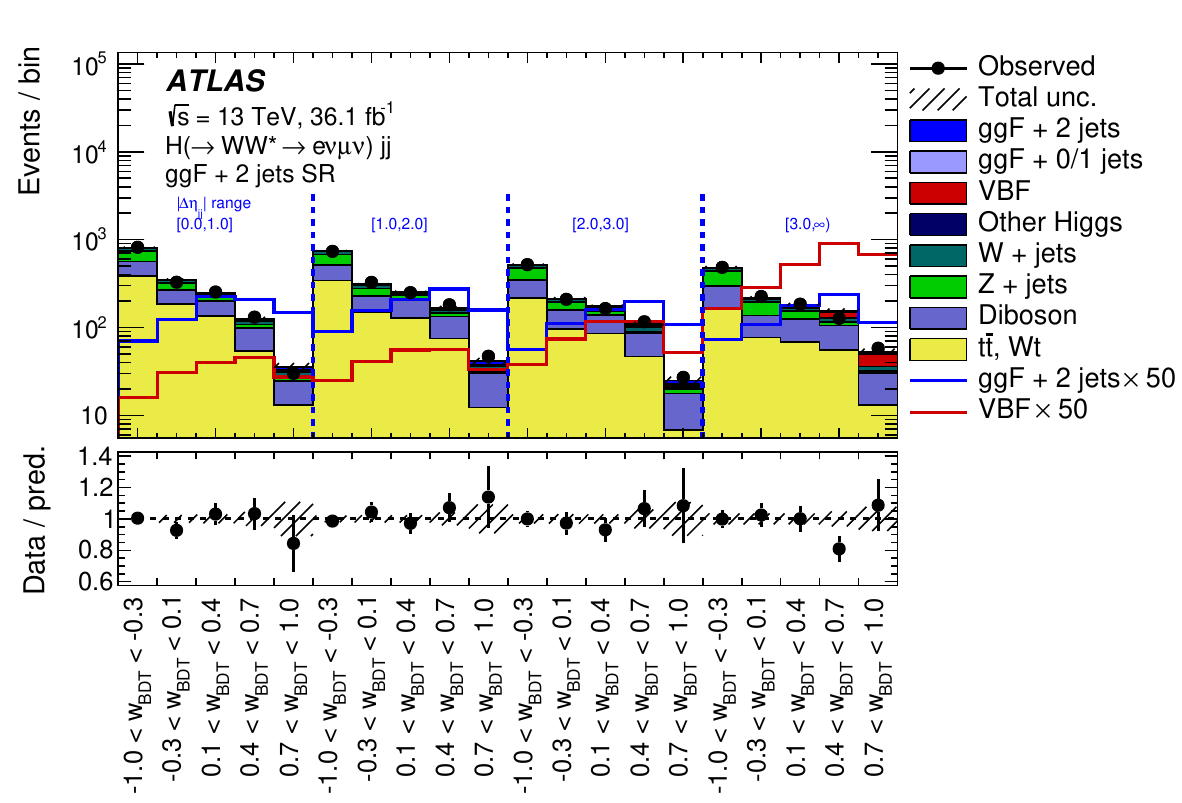}}
\par\end{centering}
\caption{Post-fit distribution of the BDT response observable presented in the four $|\Delta\eta_{jj}|$ categories of the
$\mathrm{ggF}$ + 2 jets signal region, with signal and background yields fixed from the fit for $\mu^{\mathrm{ggF}+\mathrm{2 jets}}$.
Data-to-simulation ratios are shown at the bottom of the plot. The shaded areas depict the total uncertainty. The distributions of
the $\mathrm{ggF}$ + 2 jets and VBF processes are overlaid with their respective contributions multiplied by 50.}
\label{fig:PostFitDistribution_MuFITINPUT}
\end{figure}

\begin{figure}[!h]
\noindent \begin{centering}
\subfloat{\includegraphics[width=0.6\textwidth]{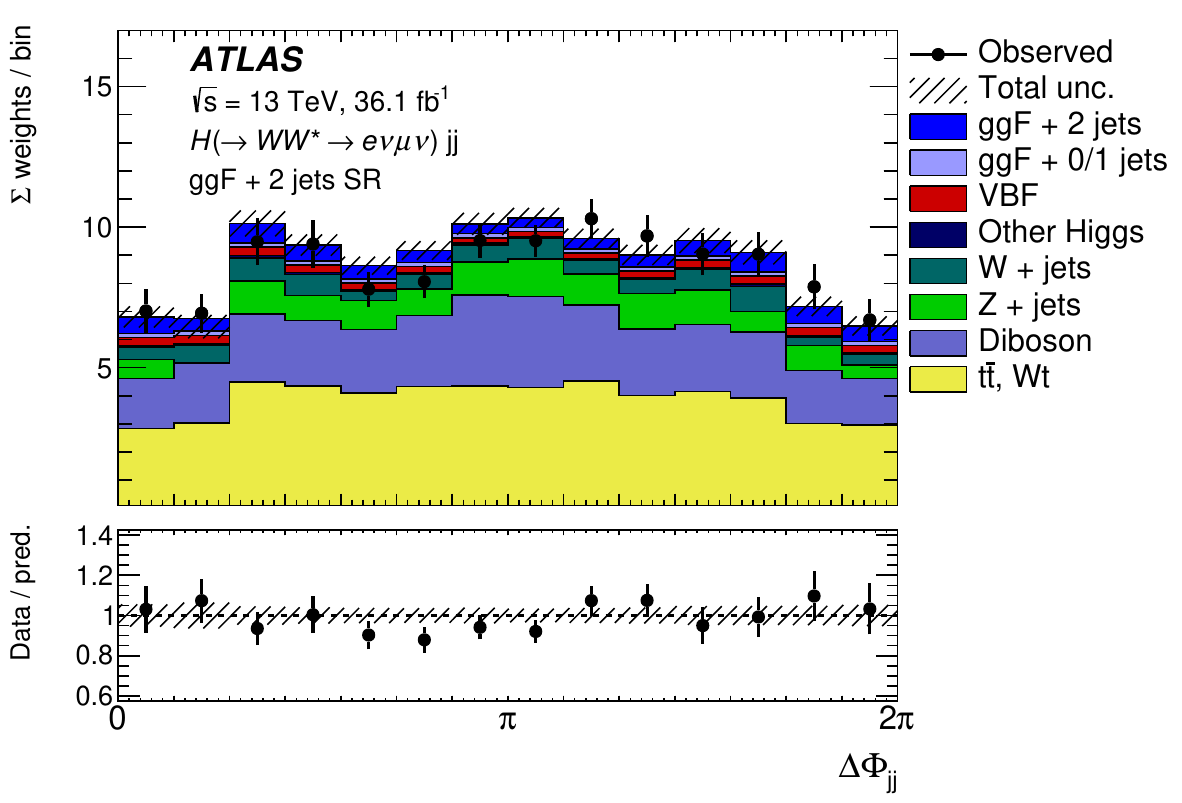}}
\par\end{centering}
\caption{The weighted $\Delta\Phi_{jj}$ post-fit distribution in the $\mathrm{ggF}$ + 2 jets signal region, with signal
and background yields fixed from the fit to $\tan(\alpha)$ using shape and rate
information. Data-to-simulation ratios are shown at the bottom of the plot.
The shaded areas depict the total uncertainty.}
\label{fig:PostFitDistribution_forTanAlphaShapeRate_DataFit_muVBF_fixed}
\end{figure}

The post-fit distribution of the inputs to the signal strength parameter $\mu^\mathrm{ggF + 2 jets}$ determination is depicted in Figure~\ref{fig:PostFitDistribution_MuFITINPUT}.
The normalised $\Delta\Phi_{jj}$ distribution in the various event categories of the $\mathrm{ggF}$ + 2 jets signal
region is depicted in Figure \ref{fig:PostFitDistribution_forTanAlphaShapeRate_DataFit_muVBF_fixed}.
Events are weighted by $\ln(1+N_\mathrm{S}/N_\mathrm{B})$, where $N_\mathrm{S}$ ($N_\mathrm{B}$) is the post-fit signal (background)
event yield for each event category. This distribution is presented for the minimum of the NLL curve from the fit
configuration that exploits both shape and rate information. The post-fit event yields in the signal region and the various control
regions are presented in Table~\ref{tab:MuFit2D_EventYieldsPostFit_unblinded}. \par
 
\begin{figure}[!h]
\centering
\subfloat[]{\includegraphics[width=0.5\textwidth]{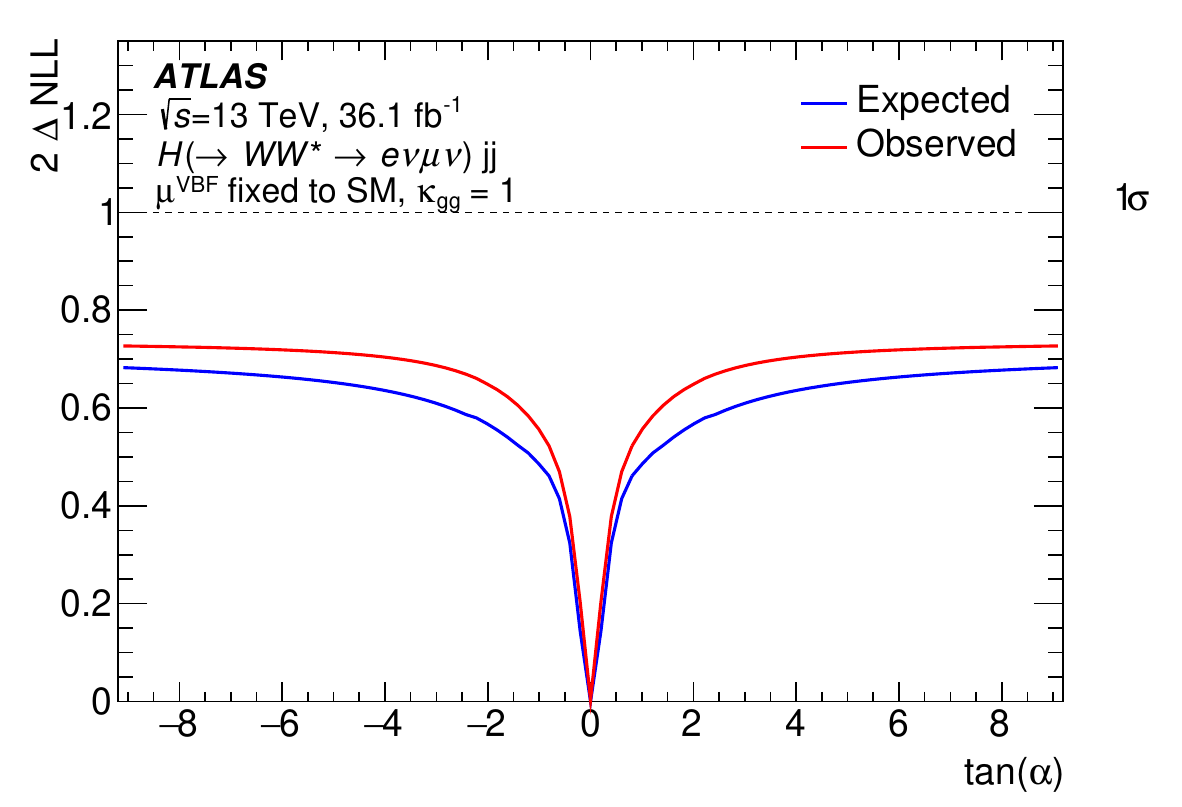}}
\subfloat[]{\includegraphics[width=0.5\textwidth]{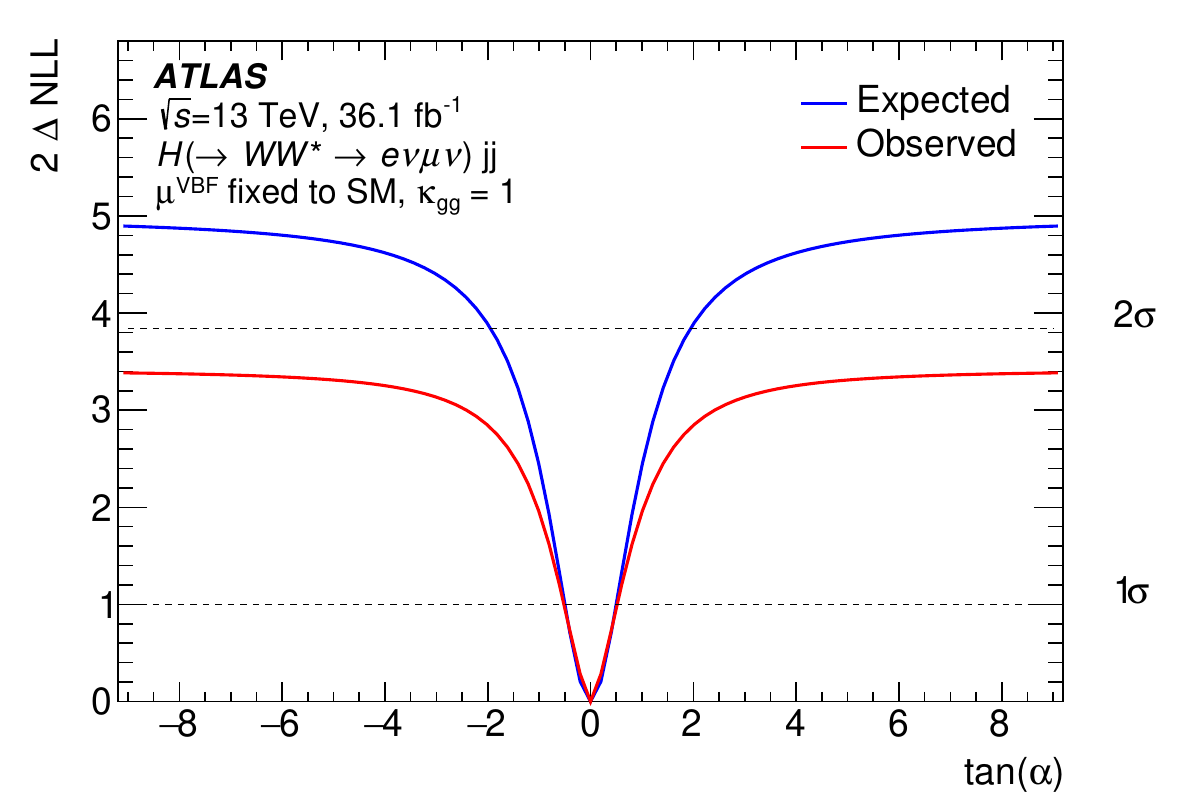}}
 
\caption{Expected and observed likelihood curves for scans (a) over $\tan(\alpha)$
where only the shape is taken into account in the fit, and (b) over $\tan(\alpha)$ when both shape and normalisation are used.}
\label{fig:tana-muvbffixed}
\end{figure}

The signal strength parameter of the $\mathrm{ggF}$ + 2 jets process is found to be $\mu^{\mathrm{ggF + 2 jets}}= 0.5 \pm 0.4 (\mathrm{stat.}) ^{+0.7}_{-0.6} (\mathrm{syst.})$,
which is consistent with the SM prediction. The observed and expected likelihood curves corresponding to scans over $\tan(\alpha)$ are
presented in Figure \ref{fig:tana-muvbffixed}. The values of the NLL are evaluated in steps
of $\Delta\left(\tan(\alpha)\right) = 0.2$, and the minima of all fits are consistent with zero, i.e.\ the
SM hypothesis. If only the shape information is used, the data are not sensitive enough
to provide $68\%$ confidence level (CL) intervals on $\tan(\alpha)$. The ML fit configuration
that uses both shape and rate information provides a best-fit value of $\tan(\alpha) = 0.0 \pm 0.4 (\mathrm{stat.}) \pm 0.3 (\mathrm{syst.})$
for both the fits to data and to Asimov data. The observed sensitivity is slightly worse than the expected sensitivity due
to a lower than expected signal yield (consistent with a signal strength value below unity). Hence, the
fit to this dataset does not provide a $95\%$ CL interval. The relative impact of the main uncertainties on $\tan(\alpha)$ is
presented in Table~\ref{tab:TanAlphaFit_ShapeRateOnData}.\par
 
\begin{figure}[!h]
\centering
\includegraphics[width=0.65\textwidth]{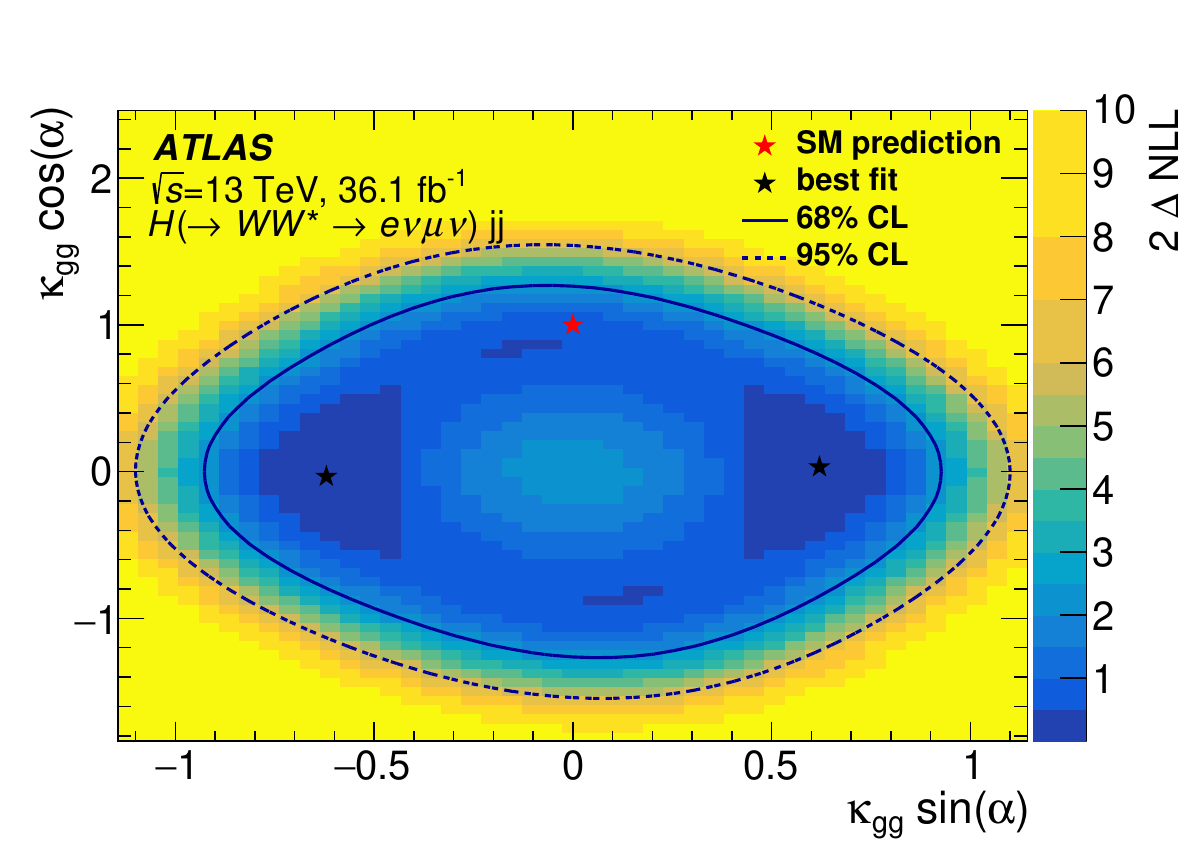}
\caption{$68\%$ and $95\%$ CL two-dimensional likelihood contours of the CP-even
and CP-odd coupling parameters $\kappa^{}_{gg}\cos(\alpha)$ and $\kappa^{}_{gg}\sin(\alpha)$.
The minima are represented by black stars, while the SM point is shown as a red star.}
\label{fig:TwoDimensionalFitOfCPevenANDCPodd}
\end{figure}

The $68\%$ and $95\%$ CL two-dimensional likelihood contours of the
simultaneous scan over the CP-even and CP-odd coupling strength parameters $\kappa^{}_{gg}\cos(\alpha)$ and
$\kappa^{}_{gg}\sin(\alpha)$ are presented in Figure \ref{fig:TwoDimensionalFitOfCPevenANDCPodd}.
Two minima are found, as the signal rate depends on the square of both couplings.
The best-fit values observed in the data are consistent with the SM predictions within the $68\%$ CL,
while $|\kappa^{}_{gg} \cos(\alpha)|$ values above $1.6$ and $|\kappa^{}_{gg}\sin(\alpha)|$ values above $1.1$ are excluded at
$95\%$ CL.\footnote{Precise measurements of the inclusive ggF cross section give tighter
constraints on the individual parameters~\cite{hcomb}, due to its dependence on
$\kappa_{gg}^2\cos^2(\alpha)$ and $\kappa_{gg}^2\sin^2(\alpha)$.} \par

\begin{table}[!h]
\caption{Post-fit event yields in the signal and control regions obtained from the study of the
signal strength parameter $\mu^\mathrm{ggF + 2 jets}$. The quoted
uncertainties include those from theoretical and experimental systematic sources and those due to
sample statistics. The fit constrains the total expected yield to the observed yield.
The diboson background is split into $WW$ and non-$WW$ contributions.}
\centering
\small
\renewcommand{\arraystretch}{1.15}
\begin{tabular}{c c c c c}
\hline
\hline
Process & Top CR &  $WW$ CR & $Z\rightarrow \tau\tau$ CR & SR \\
\hline
$\mathrm{ggF}$ + 2 jets           & $20\pm20$    & $<0.1$        & $10\pm10$    & $60\pm80$ \\
$\mathrm{ggF}$ + 0/1 jets         & $4\pm1$      & $<0.1$        & $3\pm1$      & $40\pm20$ \\
VBF                               & $8\pm1$      & $<0.1$        & $7\pm1$      & $70\pm10$ \\
Other Higgs                       & $6\pm3$  & \;\;$2\pm1$       & $20\pm10$    & $30\pm10$ \\
\hline
$t\bar{t}$, $Wt$                  & $17800\pm200$~~~~ & $3100\pm500$  & $390\pm60$~~   & $2300\pm300$~~ \\
$WW$                              & $180\pm80$~~   & $1400\pm500$  & $200\pm70$~~   & $1200\pm400$~~ \\
$Z$ + jets                        & $220\pm30$~~   & $16\pm3$      & $1960\pm70$~~~~  & $1000\pm100$~~ \\
$W$ + jets                        & $600\pm200$  & $140\pm30$    & $90\pm20$    & $390\pm80$~~ \\
Non-$WW$ dibosons                 & $40\pm30$    & $100\pm30$    & $120\pm50$~~   & $240\pm80$~~ \\
\hline
Observed                          &     $18886$   &     $4778$   & $2800$       &    $5209$  \\
\hline
\hline
\end{tabular}
\label{tab:MuFit2D_EventYieldsPostFit_unblinded}
\end{table}
 
\clearpage
 
\begin{table}[!h]
\caption{Breakdown of the main contributions to the total uncertainty in $\tan(\alpha)$ based
on the fit that exploits both shape and rate information. Individual sources of systematic
uncertainty are grouped into either the theoretical
or the experimental uncertainty. The sum in quadrature of the individual components
differs from the total uncertainty due to correlations between the components.}
\centering
\small
\renewcommand{\arraystretch}{1.15}
\begin{tabular}{l c}
\hline
\hline
Source & $\Delta \left(\tan(\alpha)\right)$ \tabularnewline
\hline
Total data statistical uncertainty & \ensuremath{0.4}~~ \tabularnewline
\;\;\;SR statistical uncertainty & \ensuremath{0.33} \tabularnewline
\;\;\;CR statistical uncertainty & \ensuremath{0.10} \tabularnewline
\hline
MC statistical uncertainty & \ensuremath{0.14} \tabularnewline
\hline
Total systematic uncertainty & \ensuremath{0.28} \tabularnewline
Theoretical uncertainty & \ensuremath{0.23} \tabularnewline
\;\;\; Top-quark bkg. & \ensuremath{0.15} \tabularnewline
\;\;\; $\mathrm{ggF}$ signal & \ensuremath{0.14} \tabularnewline
\;\;\; $WZ$, $ZZ$, $W\gamma$, $Z\gamma$ bkg. & \ensuremath{0.06} \tabularnewline
\;\;\; $WW$ bkg. & \ensuremath{0.06} \tabularnewline
\;\;\; $Z/\gamma^{*}$ bkg. & ~~\ensuremath{0.016} \tabularnewline
\;\;\; VBF bkg. & ~~\ensuremath{0.015} \tabularnewline
\hline
Experimental uncertainty & \ensuremath{0.21} \tabularnewline
\;\;\; $b$-tagging & \ensuremath{0.16} \tabularnewline
\;\;\; Modelling of pile-up  & \ensuremath{0.10} \tabularnewline
\;\;\; Jets & \ensuremath{0.07} \tabularnewline
\;\;\; Misidentified leptons  & \ensuremath{0.04} \tabularnewline
\;\;\; Luminosity & ~~\ensuremath{0.034} \tabularnewline
\hline
Total & \ensuremath{0.5}~~ \tabularnewline
\hline
\hline
\end{tabular}
\label{tab:TanAlphaFit_ShapeRateOnData}
\end{table}
 
\clearpage
\subsection{VBF category}
 
To constrain the polarisation-dependent coupling-strength scale factors in the VBF production process,
the signal region and background control regions defined in Tables~\ref{tab:SignalRegionsDef} and \ref{tab:ControlRegionsDef} are used
as inputs to the ML fit.
In the signal region, the distribution of the signed $\Delta\Phi_{jj}$ observable is used in the four
categories defined by the BDT-score intervals $[-1,0.26,0.61,0.86,1]$, following Ref.~\cite{atlas-higg-2016-07}.
The first (last) interval corresponds to the enhanced background (signal) phase-space region. No shape information
is used in the control regions.
 
Several types of fits are performed in the VBF event category with both $(a_{\mathrm{L}}, a_{\mathrm{T}})$ and $(\kappa_{VV}, \varepsilon_{VV})$ parameterisations:
\begin{itemize}
\item One-dimensional fits are performed
\begin{itemize}
\item using only the dependence of the input distribution's shape on the selected parameter of interest, with the other parameter fixed to its SM value;
\item  exploiting both shape and rate information, fixing one parameter to its SM value.
\end{itemize}
The parameters $a_{\mathrm{L}}$ and $\kappa_{VV}$ dominantly affect the total event yields because the longitudinal
polarisation vectors of the massive gauge bosons are proportional to the energy and can give large enhancements to the
total cross section. These parameters are only weakly constrained by the shape of the $\Delta\Phi_{jj}$ distribution.
The parameters $a_{\mathrm{T}}$ and $\varepsilon_{VV}$, on the other hand, affect the shape of the
$\Delta\Phi_{jj}$ distribution and can be constrained from both types of fits. The kinematic distributions of the two tagging jets are related to the intrinsic structure
of the Higgs boson production vertex and carry information about the polarisation of the fusing gauge bosons.
\item Fits to one parameter are performed with the other being profiled\footnote{A profiled parameter is fixed to its unconditional maximum likelihood value for each value of the parameter of interest (POI). Then, the POI is fixed to its conditional maximum likelihood value.}.
In these fits, both parameters may vary independently in order to probe them in a model-independent way.
\end{itemize}
 
\begin{figure}[!ht]
\noindent \begin{centering}
\includegraphics[width=0.515\textwidth]{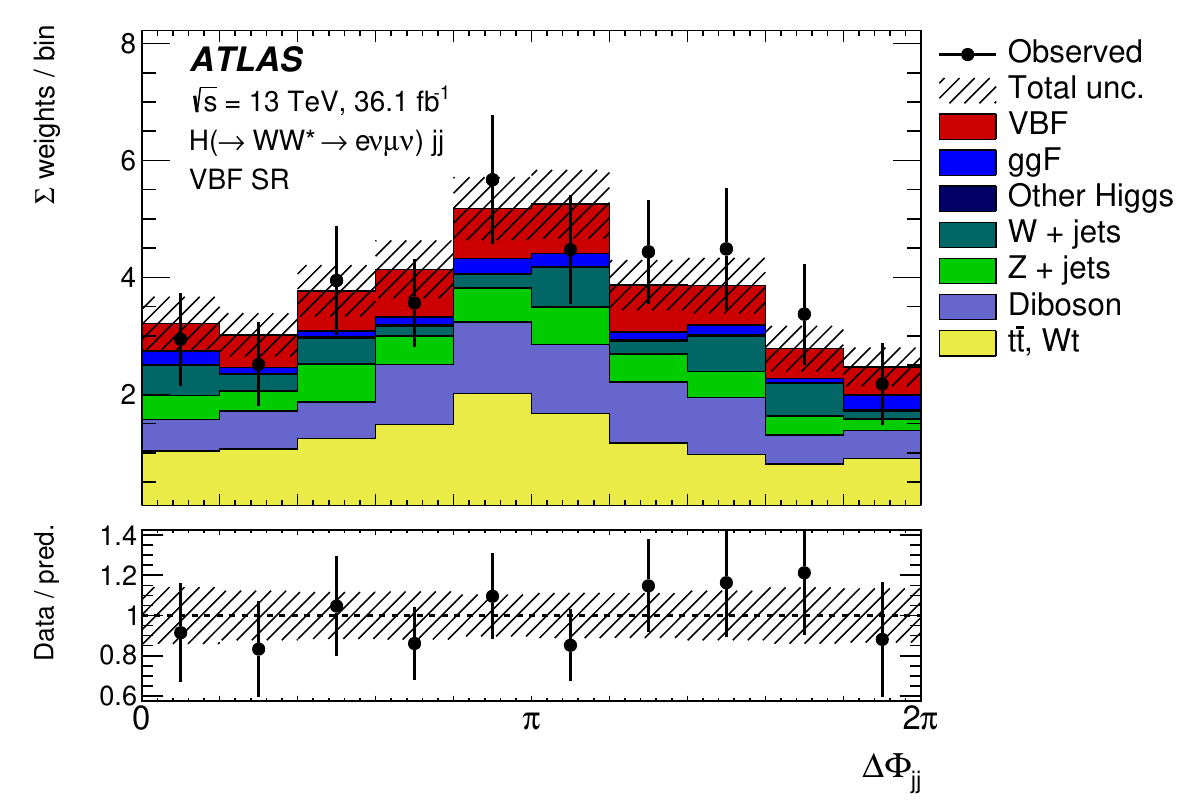}
\par\end{centering}
\caption{The weighted $\Delta\Phi_{jj}$ distribution in the VBF signal region, with signal
and background yields fixed from the fit for $\varepsilon_{VV}$ using shape and rate
information. Data-to-simulation ratios are shown at the bottom of the plot.
The shaded areas depict the total uncertainty.}
\label{fig:PostFitDistribution_VBF_weighted}
\end{figure}

\begin{figure}[!h]\centering
\subfloat[$a_\mathrm{T}$ shape-only fit with $a_{\mathrm{L}}=1$]{\includegraphics[width=0.475\textwidth]{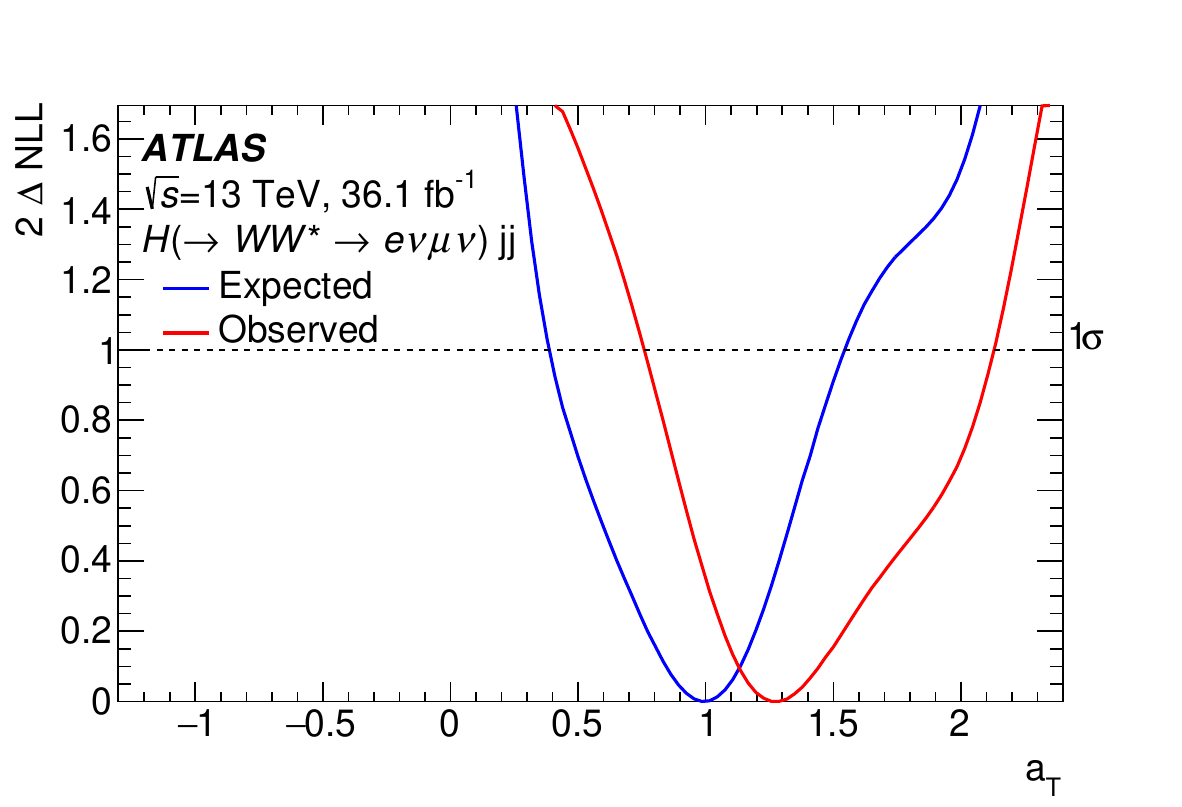}}
\subfloat[$a_\mathrm{T}$ shape+rate fit with $a_{\mathrm{L}}=1$]{\includegraphics[width=0.475\textwidth]{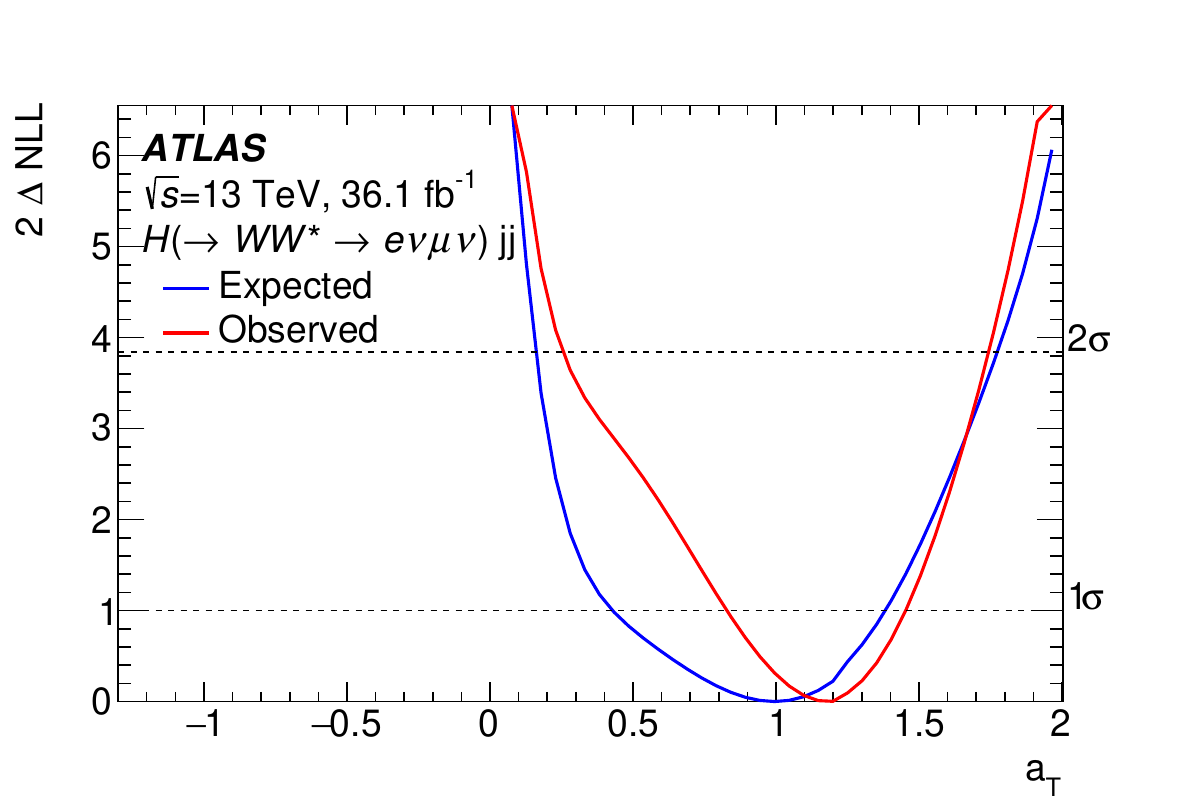}}\\
\subfloat[$a_\mathrm{L}$ shape+rate fit with $a_{\mathrm{T}}=1$]{\includegraphics[width=0.475\textwidth]{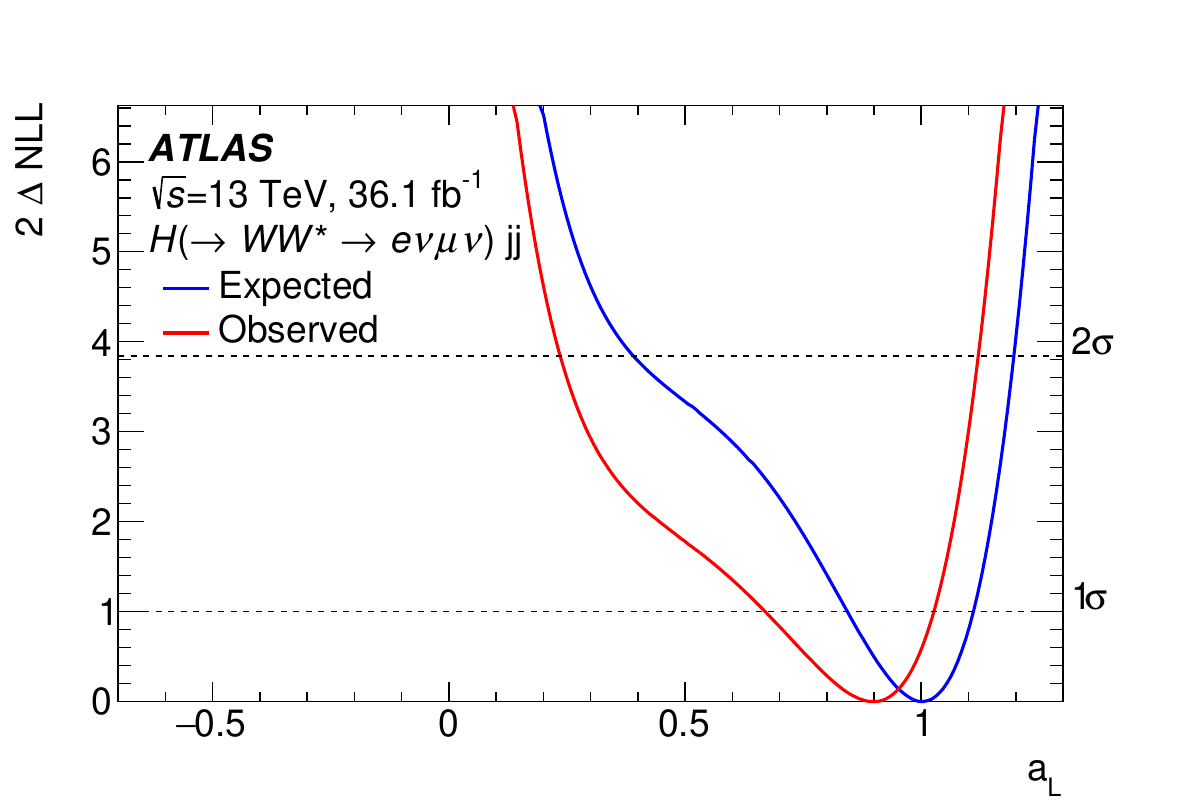}}
\caption{Likelihood scans over the transversely (a, b) and longitudinally (c) polarised couplings.
Fits using shape-only (a) and shape+rate (b, c) information are shown. All relevant experimental and
modelling systematic uncertainties are considered in the fits.}
\label{fig:likelihoodscans}
\end{figure}

Figure~\ref{fig:PostFitDistribution_VBF_weighted} depicts the weighted $\Delta\Phi_{jj}$ distribution in all categories of the VBF signal region.
Events are weighted by $\ln(1+N_\mathrm{S}/N_\mathrm{B})$ in their corresponding event category. The post-fit event
yields in the signal  and the control regions are presented in Table~\ref{tab:VBF_ALfit_EventYieldsPostFit}. The results of the likelihood scans
are given in Figure~\ref{fig:likelihoodscans}. The scan over $a_{\mathrm{L}}$ ($a_{\mathrm{T}}$) is shown in the upper (lower) panel.
Both scans have been performed in two configurations: the LLH curves shown in Figures~\ref{fig:likelihoodscans}(b) and \ref{fig:likelihoodscans}(c) are the results of the fit in which both
the shape and normalisation of the signal are taken into account, while the LLH curves in Figure~\ref{fig:likelihoodscans}(a) are obtained using only the shape information.
The largest sensitivity to $a_\mathrm{L}$ stems from the rate information.
The asymmetry of the curves results from the asymmetric behaviour of the cross section with respect to the changes in the parameter values (see Ref.~\cite{Brehmer:2014pka}).
The expected sensitivity  to $a_\mathrm{T}$ comes predominantly from the shape information, as the likelihood ratio increases only slightly when adding the
normalisation information. The resulting best-fit values and their uncertainties computed at $68\%$ CL are presented in
Table~\ref{tab:results_ALAT}. All measurements are consistent with the SM expectations.
The dominant sources of uncertainty are related to the limited data yields and to the uncertainties in modelling the top-quark and
$WW$ backgrounds. \par
 
\begin{figure}[!ht]
\noindent \begin{centering}
\subfloat[$\kappa_{VV}$ shape+rate fit with profiled $\varepsilon_{VV}$]{\includegraphics[width=0.475\textwidth]{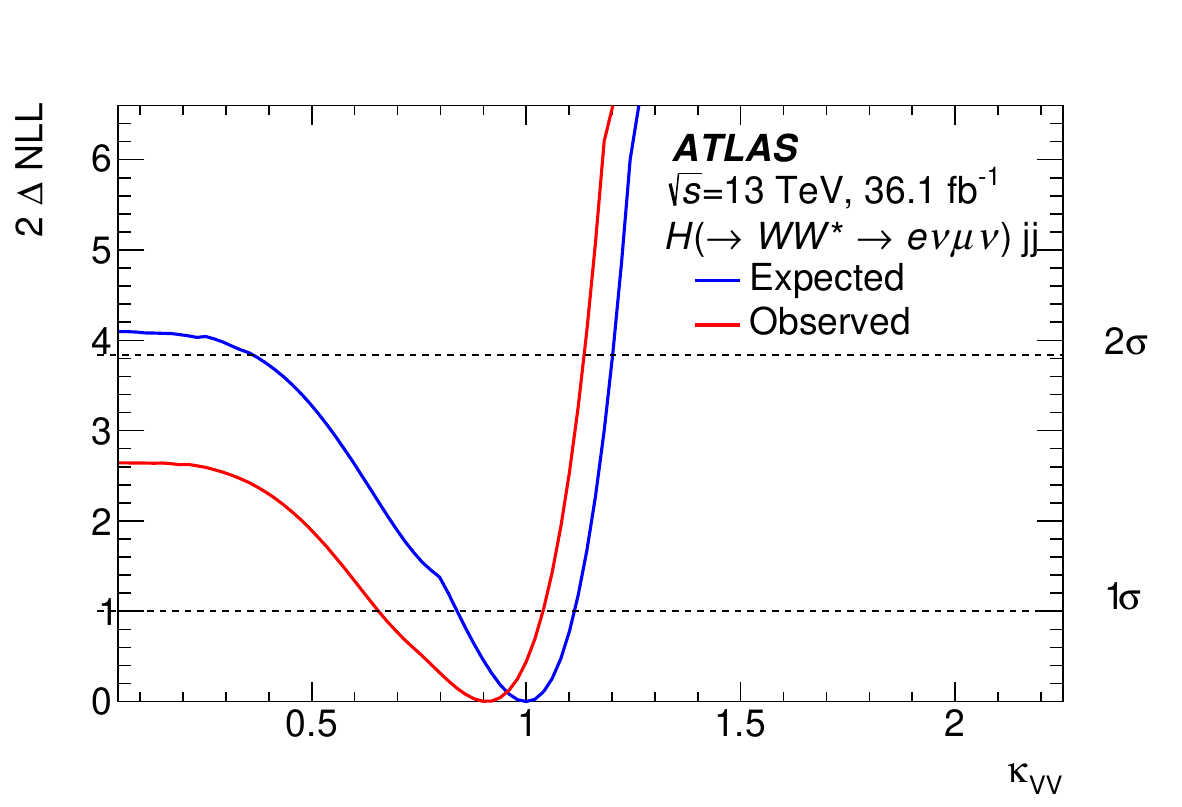}}
\subfloat[$\varepsilon_{VV}$ shape+rate fit with profiled $\kappa_{VV}$]{\includegraphics[width=0.475\textwidth]{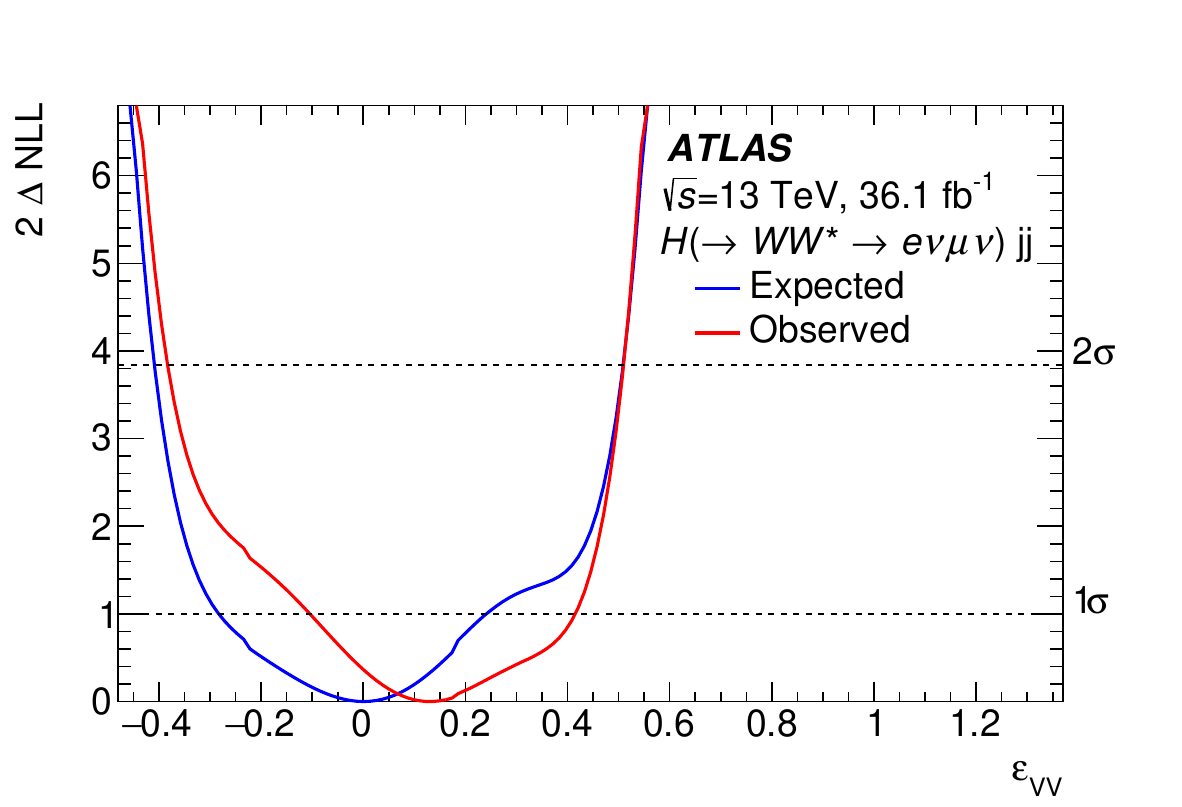}}
\par\end{centering}
\caption{Likelihood scans over (a) $\kappa_{VV}$ and (b) $\varepsilon_{VV}$, with the other parameter profiled.
The fits are performed using both shape and rate information. All relevant experimental and theoretical systematic uncertainties
are considered in the fits.}
\label{fig:POScans_unblinded2}
\end{figure}

The results of the ML scans of pseudo-observables are shown in Table~\ref{tab:results_PO}, with the uncertainty breakdown given in
Table~\ref{tab:PO_ShapeRateOnData}. Figure~\ref{fig:POScans_unblinded2} shows the expected and observed likelihood curves for scans
over one pseudo-observable with the other one profiled. Both shape and rate information are employed.
 
\begin{table}[!h]
\caption{Post-fit event yields in the signal and control regions obtained from a scan over $\varepsilon_{VV}$ exploiting both shape
and rate information. The quoted uncertainties include those from theoretical and experimental systematic sources and those
due to sample statistics. The fit constrains the total expected yields to the observed yields.
The diboson background is split into $WW$ and non-$WW$ contributions.}
\centering
\renewcommand{\arraystretch}{1.15}
\begin{tabular}{c c c c}
\hline
\hline
Process & Top CR & $Z\rightarrow \tau\tau$ CR & SR \\
\hline
VBF            & $3.2\pm2.2$  & $2.6\pm1.8$  & $34\pm22$  \\
$\mathrm{ggF}$ & $3.9\pm1.7$  & $2.4\pm1.0$  & $28\pm12$  \\
Other Higgs    & $1.5\pm 0.7$ & $6.2\pm3.1$  & $6.0\pm 3.0$  \\
\hline
$t\bar{t}$, $Wt$ & $7400\pm100$~~  & $53\pm7$~~  & $1220\pm100$~~ \\
$WW$ & $51\pm6$~~  & $21.8\pm2.9$~~  & $360\pm70$~~  \\
$Z$ + jets & $54\pm10$  & $370\pm24$~~  & $320\pm70$~~ \\
$W$ + jets & $190\pm40$~~  & $23.0\pm2.4$~~  & $115\pm27$~~ \\
Non-$WW$ dibosons & $14.3\pm1.8$~~  & $20.8\pm3.3$~~  & $83\pm11$  \\
\hline
Observed & $7668$  & $501$  & $2164$ \\
\hline
\hline
\end{tabular}
\label{tab:VBF_ALfit_EventYieldsPostFit}
\end{table}
 
\begin{table}[!h]
\centering
\caption{Best-fit values and their uncertainties as obtained from the shape-only and shape-plus-rate likelihood fits to
the Asimov dataset and to ATLAS data. Results of both the shape-only and shape+rate fits for $a_{\mathrm{L}}$  and $a_{\mathrm{T}}$
are shown. Results of fits to one parameter with the other one fixed or profiled are presented.}
\renewcommand{\arraystretch}{1.5}
\begin{tabular}{c c c }
\toprule
Type                      &   Expected  & Observed   \\
\midrule
$a_{\mathrm{T}}$ shape-only fit ($a_{\mathrm{L}} = 1$)   &   $1.0\pm 0.5 $(stat.)$^{+0.3}_{-0.4} $(syst.)  &  $1.3^{+0.8}_{-0.4}$(stat.)$^{+0.3}_{-0.2}$(syst.)    \\
\midrule
$a_{\mathrm{L}}$ shape + rate fit ($a_{\mathrm{T}} = 1$) &   $1.00^{+0.08}_{-0.10}$(stat.)$^{+0.07}_{-0.13}$(syst.)  & $0.90^{+0.09}_{-0.13}$(stat.)$^{+0.08}_{-0.18}$(syst.)  \\
$a_{\mathrm{T}}$ shape + rate fit ($a_{\mathrm{L}} = 1$) &   $1.00^{+0.36}_{-0.49}$(stat.)$^{+0.19}_{-0.27}$(syst.)  & $1.19^{+0.27}_{-0.32}$(stat.)$^{+0.12}_{-0.14}$(syst.)   \\
\midrule
$a_{\mathrm{L}}$ shape + rate fit ($a_{\mathrm{T}}$ profiled) &   $1.00^{+0.08}_{-0.10}$(stat.)$^{+0.08}_{-0.13}$(syst.)  &  $0.91^{+0.10}_{-0.18}$(stat.)$^{+0.09}_{-0.17}$(syst.)  \\
$a_{\mathrm{T}}$ shape + rate fit ($a_{\mathrm{L}}$ profiled) &   $1.0^{+0.4}_{-0.5}$(stat.)$^{+0.2}_{-0.4}$(syst.)    &  $1.2 \pm 0.4 $(stat.)$ ^{+0.2}_{-0.3} $(syst.)    \\
\bottomrule
\end{tabular}
\label{tab:results_ALAT}
\end{table}
 
\clearpage
 
\begin{table}[!ht]
\caption{
Best-fit values and their uncertainties as obtained from the shape-only and shape-plus-rate likelihood fits to
the Asimov dataset and to ATLAS data. Results of both shape-only and shape+rate fits for $\varepsilon_{VV}$  and $\kappa_{VV}$
are shown. Results of fits to one parameter with the other one fixed or profiled are presented.}
\centering
\renewcommand{\arraystretch}{1.5}
\begin{tabular}{c c c }
\toprule
Type                               &   Expected                      & Observed   \\
\midrule
$\varepsilon_{VV}$ shape-only fit ($\kappa_{VV} = 1$     )      &   $0.00^{+0.23}_{-0.25}$(stat.)$^{+0.14}_{-0.17}$(syst.)    & $0.14^{+0.39}_{-0.22}$(stat.)$^{+0.16}_{-0.12}$(syst.)   \\
\midrule
$\kappa_{VV}$      shape + rate fit ($\varepsilon_{VV} = 0$)    &   $1.00^{+0.08}_{-0.10}$(stat.)$^{+0.08}_{-0.13}$(syst.)   & $0.91^{+0.09}_{-0.12}$(stat.)$^{+0.09}_{-0.18}$(syst.)   \\
$\varepsilon_{VV}$ shape + rate fit ($\kappa_{VV} = 1$     )    &  $0.00^{+0.18}_{ -0.24}$(stat.)$^{+0.08}_{-0.13}$(syst.)     &  $0.09^{+0.13}_{-0.16}$(stat.)$^{+0.06}_{-0.07}$(syst.)   \\
\midrule
$\kappa_{VV}$      shape + rate fit ($\varepsilon_{VV}$ profiled) &   $1.00^{+0.08}_{-0.10}$(stat.)$^{+0.08}_{-0.13}$(syst.)   & $ 0.91^{+0.10}_{-0.18}$(stat.)$^{+0.09}_{-0.17}$(syst.)   \\
$\varepsilon_{VV}$ shape + rate fit ($\kappa_{VV}$      profiled) &   $0.00^{+0.22}_{-0.24}$(stat.)$^{+0.11}_{-0.15}$(syst.)  &  $0.13^{+0.28}_{-0.20}$(stat.)$^{+0.08}_{-0.10}$(syst.)    \\
\bottomrule
\end{tabular}
\label{tab:results_PO}
\end{table}

\begin{table}[!h]
\caption{The contributions of the leading individual systematic uncertainties together with
the data statistical uncertainties, in the one-dimensional fit for the pseudo-observables
$\kappa_{VV}$ (a) and $\varepsilon_{VV}$ (b) for electroweak-boson polarisation
in the $\mathrm{VBF}$ $H\to WW$ channel. Both the shape and rate information is exploited in the fit.
The theoretical and experimental uncertainties are subdivided further into categories.}
\centering
\subfloat[$\kappa_{VV}$ fit, $\varepsilon_{VV}=0$]{
\renewcommand{\arraystretch}{1.15}
\begin{tabular}{l c}
\hline
\hline
Source & $\Delta\kappa_{VV}$ \tabularnewline
\hline
Total data statistical uncertainty & \ensuremath{0.11}~~  \tabularnewline
\;\;\; SR data statistical uncertainty & \ensuremath{0.10}~~ \tabularnewline
\;\;\; CR data statistical uncertainty & \ensuremath{0.019} \tabularnewline
\hline
MC statistical uncertainty                  & \ensuremath{0.035} \tabularnewline
\hline
Total systematic uncertainty       & \ensuremath{0.12}~~ \tabularnewline
Theoretical uncertainty        & \ensuremath{0.10}~~ \tabularnewline
\;\;\; Top-quark bkg.          & \ensuremath{0.072} \tabularnewline
\;\;\; $WW$ bkg.               & \ensuremath{0.062} \tabularnewline
\;\;\; ggF bkg.                & \ensuremath{0.033} \tabularnewline
\;\;\; $Z/\gamma^{*}$ bkg.     & \ensuremath{0.017} \tabularnewline
\;\;\; VBF signal              & \ensuremath{0.019} \tabularnewline
\hline
Experimental uncertainty       & \ensuremath{0.050} \tabularnewline
\;\;\; Jet                     & \ensuremath{0.026} \tabularnewline
\;\;\; $b$-tagging             & \ensuremath{0.014} \tabularnewline
\;\;\; Luminosity              & \ensuremath{0.011} \tabularnewline
\;\;\; Misidentified leptons   & \ensuremath{0.007} \tabularnewline
\hline
Total & \ensuremath{0.17}~~ \tabularnewline
\hline
\hline
\end{tabular}}
\subfloat[$\varepsilon_{VV}$ fit, $\kappa_{VV}=1$]{
\renewcommand{\arraystretch}{1.15}
\begin{tabular}{l c }
\hline
\hline
Source & $\Delta\varepsilon_{VV}$ \tabularnewline
\hline
Total data statistical uncertainty  & \ensuremath{0.14}~~ \tabularnewline
\;\;\; SR data statistical uncertainty & \ensuremath{0.14}~~ \tabularnewline
\;\;\; CR data statistical uncertainty & \ensuremath{0.011} \tabularnewline
\hline
MC statistical uncertainty                  & \ensuremath{0.036} \tabularnewline
\hline
Total systematic uncertainty       & \ensuremath{0.056} \tabularnewline
Theoretical uncertainty        & \ensuremath{0.050} \tabularnewline
\;\;\; Top-quark bkg.          & \ensuremath{0.039} \tabularnewline
\;\;\; $WW$ bkg.               & \ensuremath{0.036} \tabularnewline
\;\;\; ggF bkg.                & \ensuremath{0.013} \tabularnewline
\;\;\; $Z/\gamma^{*}$ bkg.     & \ensuremath{0.012} \tabularnewline
\;\;\; VBF signal              & \ensuremath{0.010} \tabularnewline
\hline
Experimental uncertainty       & \ensuremath{0.024} \tabularnewline
\;\;\; Modelling of pile-up    & \ensuremath{0.022} \tabularnewline
\;\;\; Jet                     & \ensuremath{0.018} \tabularnewline
\;\;\; Misidentified leptons   & \ensuremath{0.010} \tabularnewline
\;\;\; $b$-tagging             & \ensuremath{0.010} \tabularnewline
\hline
Total & \ensuremath{0.16}~~ \tabularnewline
\hline
\hline
\end{tabular}}
\label{tab:PO_ShapeRateOnData}
\end{table}
 
\section{Conclusion}\label{sec:conclusion}
This article presents constraints on the CP structure of gluon--gluon fusion Higgs boson production and on the polarisations of the vector bosons in the $HVV$ coupling.
The results are obtained using $H(\rightarrow WW^*\rightarrow e\nu\mu\nu) jj$ final states
in data corresponding to 36.1\;fb$^{-1}$ of $\sqrt{s} = 13\;\TeV$ proton--proton collisions
recorded by the ATLAS detector at the LHC in 2015--2016.
Total event yields as well as shapes of selected kinematical distributions
in signal and control regions are exploited. \par
 
The signal strength parameter for the $\mathrm{ggF}$ + 2 jets Higgs boson production mode is found to be
$\mu^{\mathrm{ggF + 2 jets}} = 0.5 \pm 0.4 (\mathrm{stat.}) ^{+0.7}_{-0.6} (\mathrm{syst.})$.
The mixing angle for CP-even and CP-odd contributions to the effective Higgs--gluon interaction is determined
to be $\tan(\alpha) = 0.0 \pm 0.4 (\mathrm{stat.}) \pm 0.3 (\mathrm{syst.})$
using both shape and rate information. These results are obtained with novel analysis techniques and complement
existing studies of the CP properties of the Higgs boson performed using $ttH$ production.
 
The reported results for the VBF Higgs boson production mode include constraints on coupling-strength form factors to
longitudinally and transversely polarised $W$ and $Z$ bosons. It is the first such measurement of the Higgs couplings performed.
In one-dimensional ML fits (where the other parameter was set to its SM value) shape information is sufficient to constrain the coupling to transversely
polarised bosons, $a_\mathrm{T}$, while to constrain $a_\mathrm{L}$ the information about the rates significantly improves the sensitivity.
Profiling the other coupling-strength scale factor results in:
$a_\mathrm{L}=0.91^{+0.10}_{-0.18}$(stat.)$^{+0.09}_{-0.17}$(syst.) and  $a_{\mathrm{T}}=1.2 \pm 0.4$(stat.)$^{+0.2}_{-0.3} $(syst.),
while $a_{\mathrm{L}}=1.00^{+0.08}_{-0.10}$(stat.)$^{+0.08}_{-0.13}$(syst.) and $a_{\mathrm{T}}=1.0^{+0.4}_{-0.5}$(stat.)$^{+0.2}_{-0.4}$(syst.) are expected
in the SM.
 
With an approximate mapping to pseudo-observables the following constraints are obtained in a fit where the other parameter is profiled:
$\kappa_{VV}= 0.91^{+0.10}_{-0.18}$(stat.)$^{+0.09}_{-0.17}$(syst.) and $\epsilon_{VV} =0.13^{+0.28}_{-0.20}$ (stat.)$^{+0.08}_{-0.10}$(syst.),
while $\kappa_{VV}=1.00^{+0.08}_{-0.10}$(stat.)$^{+0.08}_{-0.13}$(syst.) and $\epsilon_{VV}=0.00^{+0.22}_{-0.24}$ (stat.)$^{+0.11}_{-0.15}$(syst.) are expected.
In this parameterisation the sensitivity to the on-shell coupling $\kappa_{VV}$ is affected by the event yields, while the off-shell coupling $\varepsilon_{VV}$ is
sensitive to both shape and rate information.
All measurements are consistent with the expectations for the SM Higgs boson.
\par
 
\FloatBarrier
\section*{Acknowledgements}
 

We thank CERN for the very successful operation of the LHC, as well as the
support staff from our institutions without whom ATLAS could not be
operated efficiently.
 
We acknowledge the support of
ANPCyT, Argentina;
YerPhI, Armenia;
ARC, Australia;
BMWFW and FWF, Austria;
ANAS, Azerbaijan;
SSTC, Belarus;
CNPq and FAPESP, Brazil;
NSERC, NRC and CFI, Canada;
CERN;
ANID, Chile;
CAS, MOST and NSFC, China;
Minciencias, Colombia;
MEYS CR, Czech Republic;
DNRF and DNSRC, Denmark;
IN2P3-CNRS and CEA-DRF/IRFU, France;
SRNSFG, Georgia;
BMBF, HGF and MPG, Germany;
GSRI, Greece;
RGC and Hong Kong SAR, China;
ISF and Benoziyo Center, Israel;
INFN, Italy;
MEXT and JSPS, Japan;
CNRST, Morocco;
NWO, Netherlands;
RCN, Norway;
MEiN, Poland;
FCT, Portugal;
MNE/IFA, Romania;
JINR;
MES of Russia and NRC KI, Russian Federation;
MESTD, Serbia;
MSSR, Slovakia;
ARRS and MIZ\v{S}, Slovenia;
DSI/NRF, South Africa;
MICINN, Spain;
SRC and Wallenberg Foundation, Sweden;
SERI, SNSF and Cantons of Bern and Geneva, Switzerland;
MOST, Taiwan;
TAEK, Turkey;
STFC, United Kingdom;
DOE and NSF, United States of America.
In addition, individual groups and members have received support from
BCKDF, CANARIE, Compute Canada and CRC, Canada;
COST, ERC, ERDF, Horizon 2020 and Marie Sk{\l}odowska-Curie Actions, European Union;
Investissements d'Avenir Labex, Investissements d'Avenir Idex and ANR, France;
DFG and AvH Foundation, Germany;
Herakleitos, Thales and Aristeia programmes co-financed by EU-ESF and the Greek NSRF, Greece;
BSF-NSF and GIF, Israel;
Norwegian Financial Mechanism 2014-2021, Norway;
NCN and NAWA, Poland;
La Caixa Banking Foundation, CERCA Programme Generalitat de Catalunya and PROMETEO and GenT Programmes Generalitat Valenciana, Spain;
G\"{o}ran Gustafssons Stiftelse, Sweden;
The Royal Society and Leverhulme Trust, United Kingdom.
 
The crucial computing support from all WLCG partners is acknowledged gratefully, in particular from CERN, the ATLAS Tier-1 facilities at TRIUMF (Canada), NDGF (Denmark, Norway, Sweden), CC-IN2P3 (France), KIT/GridKA (Germany), INFN-CNAF (Italy), NL-T1 (Netherlands), PIC (Spain), ASGC (Taiwan), RAL (UK) and BNL (USA), the Tier-2 facilities worldwide and large non-WLCG resource providers. Major contributors of computing resources are listed in Ref.~\cite{ATL-SOFT-PUB-2021-003}.
 

\clearpage
 
\printbibliography
\clearpage

\clearpage \input{atlas_authlist}

\end{document}

%% file: atlas_authlist.tex
 
\begin{flushleft}
\hypersetup{urlcolor=black}
{\Large The ATLAS Collaboration}

\bigskip

\AtlasOrcid[0000-0002-6665-4934]{G.~Aad}$^\textrm{\scriptsize 101}$,    
\AtlasOrcid[0000-0002-5888-2734]{B.~Abbott}$^\textrm{\scriptsize 127}$,    
\AtlasOrcid[0000-0002-7248-3203]{D.C.~Abbott}$^\textrm{\scriptsize 102}$,    
\AtlasOrcid[0000-0002-2788-3822]{A.~Abed~Abud}$^\textrm{\scriptsize 36}$,    
\AtlasOrcid[0000-0002-1002-1652]{K.~Abeling}$^\textrm{\scriptsize 53}$,    
\AtlasOrcid[0000-0002-2987-4006]{D.K.~Abhayasinghe}$^\textrm{\scriptsize 93}$,    
\AtlasOrcid[0000-0002-8496-9294]{S.H.~Abidi}$^\textrm{\scriptsize 29}$,    
\AtlasOrcid[0000-0002-8279-9324]{O.S.~AbouZeid}$^\textrm{\scriptsize 40}$,    
\AtlasOrcid{N.L.~Abraham}$^\textrm{\scriptsize 155}$,    
\AtlasOrcid[0000-0001-5329-6640]{H.~Abramowicz}$^\textrm{\scriptsize 160}$
% ...
\bigskip
\\

% ...
$^{29}$Physics Department, Brookhaven National Laboratory, Upton NY; United States of America.\\
% ...
$^{36}$CERN, Geneva; Switzerland.\\
% ...
$^{40}$Niels Bohr Institute, University of Copenhagen, Copenhagen; Denmark.\\
% ...
$^{53}$II. Physikalisches Institut, Georg-August-Universit\"{a}t G\"ottingen, G\"ottingen; Germany.\\
% ...
$^{93}$Department of Physics, Royal Holloway University of London, Egham; United Kingdom.\\
% ...
$^{101}$CPPM, Aix-Marseille Universit\'e, CNRS/IN2P3, Marseille; France.\\
$^{102}$Department of Physics, University of Massachusetts, Amherst MA; United States of America.\\
% ...
$^{127}$Homer L. Dodge Department of Physics and Astronomy, University of Oklahoma, Norman OK; United States of America.\\
% ...
$^{155}$Department of Physics and Astronomy, University of Sussex, Brighton; United Kingdom.\\
% ...
$^{160}$Raymond and Beverly Sackler School of Physics and Astronomy, Tel Aviv University, Tel Aviv; Israel.\\
% ...

\end{flushleft}
